%% file: pXi_v4f.tex
\documentclass[10pt,aps,twocolumn,prc,superscriptaddress,showpacs,nofootinbib,noshowkeys,floatfix,preprintnumbers]{revtex4-2}
\usepackage[dvipdfmx]{graphicx}
\usepackage{graphicx} 
\usepackage[usenames]{color}
\usepackage{amsmath,amssymb}
\usepackage{multirow}
\usepackage{longtable}
\usepackage[normalem]{ulem}
\usepackage{epstopdf}
\usepackage{times}
\usepackage[normalem]{ulem}  
\allowdisplaybreaks[4]

\renewcommand\sout{\bgroup \color{red} \ULdepth=-.5ex \ULset}
\newcommand{\physdim}[1]{\hspace{1ex} \mathrm{#1}}
\usepackage{bm}
\newcommand{\hyphen}{\mathchar`-}

\begin{document}
\title{Femtoscopic study of coupled-channel $N\Xi$ and $\Lambda\Lambda$ interactions}
\date{\today}
\author{Y. Kamiya}
\affiliation{
CAS Key Laboratory of Theoretical Physics, Institute of Theoretical Physics,
Chinese Academy of Sciences, 
Beijing 100190, China}
\affiliation{RIKEN Interdisciplinary Theoretical and Mathematical Science Program (iTHEMS), Wako 351-0198, Japan}
\author{K. Sasaki}
\affiliation{Division of Scientific Information and Public Policy, Center for Infectious Disease Education and Research (CiDER), Osaka University, Suita 565-0871, Japan}
\author{T. Fukui}
\affiliation{RIKEN Nishina Center, Wako 351-0198, Japan}
\author{T. Hyodo}
\affiliation{Department of Physics, Tokyo Metropolitan University, Hachioji 192-0397, Japan}
\affiliation{RIKEN Interdisciplinary Theoretical and Mathematical Science Program (iTHEMS), Wako 351-0198, Japan}
\author{K. Morita}
\affiliation{National Institutes for Quantum and Radiological Science and Technology, Rokkasho Fusion Institute, Rokkasho 039-3212, Japan}
\affiliation{RIKEN Nishina Center, Wako 351-0198, Japan}
\author{K. Ogata}
\affiliation{Research Center for Nuclear Physics, Osaka University, Ibaraki 567-0047, Japan}
\affiliation{Department of Physics, Osaka City University, Osaka 558-8585, Japan}
\affiliation{Nambu Yoichiro Institute of Theoretical and Experimental Physics, Osaka City University, Osaka 558-8585, Japan}

\author{A. Ohnishi}
\affiliation{Yukawa Institute for Theoretical Physics, Kyoto University, Kyoto 606-8502, Japan}
\author{T. Hatsuda}
\affiliation{RIKEN Interdisciplinary Theoretical and Mathematical Science Program (iTHEMS), Wako 351-0198, Japan}

\preprint{YITP-21-79, RIKEN-iTHEMS-Report-21, NITEP 116}

\begin{abstract}
The momentum correlation functions of $S=-2$ baryon pairs
($p\Xi^-$ and $\Lambda\Lambda$) produced in high-energy $pp$ and $p$A collisions are investigated
 on the basis of the coupled-channel formalism.
The strong interaction is described by the coupled-channel HAL QCD potential obtained by the lattice QCD simulations near physical quark masses,
while the hadronic source function is taken to be a static Gaussian form.  
The  coupled-channel effect, the threshold difference, the realistic strong interaction, and the Coulomb interaction  are fully taken into account for the first time
 in the femtoscopic analysis of baryon-baryon correlations.
 The characteristic features of the experimental data for the  $p\Xi^-$ and $\Lambda\Lambda$ pairs at LHC are  
 reproduced  quantitatively with a suitable choice of  non-femtoscopic parameters and the source size. The agreement between theory and
 experiment  indicates that  the $N\Xi$ ($\Lambda \Lambda$) interaction  is moderately (weakly) attractive without having a quasi-bound (bound) state.
\end{abstract}
\maketitle

\section{Introduction}

Dibaryons  in the  strangeness $S = -2$ sector have long been 
 attracted theoretical and experimental attention \cite{Oka:1988yq,Gal:2015rev,Clement:2016vnl}. 
  Among others,  the 
 $H(uuddss)$ dibaryon with ($I, J^\pi)=(0, 0^+)$ was suggested to be 
  a possible bound state below the $\Lambda\Lambda$ threshold  ~\cite{Jaffe:1976yi}. 
   However, discovery of the double $\Lambda$ hypernuclei~\cite{Takahashi:2001nm,Nakazawa:2010zzb}
 ruled out the deeply bound $H$ with the binding energy of $B_H > 6.91~\mathrm{MeV}$.
 Also, recent femtoscopic studies disfavor the existence of a bound state
below the $\Lambda\Lambda$ threshold~\cite{Acharya:2018gyz,Acharya:2019yvb,Morita:2014kza}.
 Moreover, the latest (2+1)-flavor lattice QCD simulations  
  indicate that there is no bound state below $\Lambda\Lambda$~\cite{Sasaki:2019qnh}.

Hence the current interest in the $S=-2$ dibaryons 
is shifted to the region around the $N\Xi$ threshold in the ($I, J^\pi)=(0, 0^+)$ channel
 where the   $N\Xi$ interaction is moderately attractive as indicated
  by  the existence of the  $\Xi$-hypernucleus ${}^{15}_\Xi\mathrm{C}$~\cite{Nakazawa:2015joa,Hayakawa:2020oam,Yoshimoto:2021ljs},
   by the  femtoscopic studies of the $N\Xi$ interaction \cite{Acharya:2019sms,Acharya:2020asf}, by  the chiral effective field theory calculation 
\cite{Haidenbauer:2015zqb,Li:2018tbt},  and    by the lattice QCD simulation~\cite{Sasaki:2019qnh}.
 Therefore,  it is of great importance to make a quantitative comparison between theoretical analysis and the experimental data
 with the $N\Xi$-$\Lambda\Lambda$ coupled channel framework 
 and  the state-of-the-art baryon-baryon  interactions (e.g. \cite{HALQCD:2020,Petschauer:2020}).
  Such studies are also crucial  for  identifying the role of $\Lambda$ and $\Xi^{-}$ in neutron star matter at several times the nuclear matter density
   in relation to  the so-called  hyperon puzzle in neutron star structure originally pointed out in Ref.~\cite{Nishizaki:2002ih},
    as well as to the observed constraints on  mass and radius  of neutron stars \cite{Lattimer:2021}.

It has been known that  the correlation function in high-energy collisions is sensitive to the interaction
when the absolute value of the scattering length ($a_0$) is comparable to
or larger than the emission source size $R$ of hadronic pairs, where  
 $R\simeq1{\hyphen}5~\mathrm{fm}$ depending on the reactions ($pp$, $p$A or AA)
~\cite{%
Morita:2014kza
,Ohnishi:2016elb
,Morita:2016auo
,Hatsuda:2017uxk
,Mihaylov:2018rva
,Haidenbauer:2018jvl
,Morita:2019rph
,Kamiya:2019uiw
}.
It has been also argued that the source size dependence
of the correlation function is useful to deduce the existence or non-existence 
of  hadronic bound states~\cite{Morita:2019rph,Kamiya:2019uiw}.

In the present paper, we focus on the momentum correlations of  $N\Xi$  and $\Lambda\Lambda$ in $pp$ and $p$A collisions.
  Recent experimental measurements of such correlations  have opened a new way to probe the hyperon 
 interactions which are not accessible in the standard scattering experiments \cite{Fabbietti:2020bfg}.
Theoretically, the correlation function can be  described by the convolution of the source function and the relative wave function
in the pair rest frame~\cite{%
Koonin:1977fh%
,Lednicky:1981su
,Pratt:1986cc,Anchishkin:1997tb%
,Lednicky:1998r}.

We consider the  coupled-channel formalism
 ($p\Xi^-$-$n\Xi^0$-$\Lambda\Lambda$ for $J=0$ and  $p\Xi^-$-$n\Xi^0$ for $J=1$)
     with the latest HAL QCD coupled-channel potential in the $s$-wave 
     obtained from the (2+1)-flavor lattice QCD 
        simulations at almost physical quark masses~\cite{Sasaki:2019qnh}.
  The threshold differences  and the  Coulomb interaction are taken into account simultaneously.
    For the source function in $pp$ and $p$A reactions,
  we take  a  static and spherically symmetric Gaussian form with a source size $R$.
   Our theoretical calculations are then compared with the experimental data
  of  $p\Xi^-$  and $\Lambda\Lambda$  correlation functions  in $pp$ and $p$A collisions at LHC~\cite{Acharya:2018gyz,Acharya:2019sms,Acharya:2019yvb,Acharya:2020asf}.
  Similar analysis with all ingredients (coupled channel,  threshold difference,  realistic strong interaction, and Coulomb interaction)  
    was recently performed for the $S=-1$ meson-baryon system ($\bar{K}N$-$\pi\Sigma$-$\pi\Lambda$) for the first time \cite{Kamiya:2019uiw}.

This article is organized as follows.
In Sect.~\ref{sec:HAL QCD}, we briefly review
the $S=-2$ baryon-baryon potential from lattice QCD calculations.
In Sect.~\ref{sec:corr_model},
the theoretical framework to calculate the $p\Xi^-$ and $\Lambda\Lambda$ correlation functions
 in the coupled-channel framework is discussed in detail.
In Sect.~\ref{sec:comp_exp}, we show the determination of the phenomenological
 parameters  from the experimental data at LHC on the basis of the formalism in the previous section.
In Sect.~\ref{sec:results},  our theoretical results of 
 $p\Xi^-$ and $\Lambda\Lambda$ correlation functions and the experimental data are compared.
 Section~\ref{sec:summary} is devoted to summary and concluding remarks.
The low energy scattering parameters from a modified HAL QCD  potential, the location of the virtual pole near the $N\Xi$ threshold,
 and an analytic model of the correlation function with Gamow factor are discussed in Appendix A, B, and C, respectively.

\section{$S=-2$ coupled-channel potential from Lattice QCD}\label{sec:HAL QCD}

\begin{figure*}[t]
	\begin{center}
		\begin{minipage}{1\hsize}
			\includegraphics[width=0.3\textwidth]{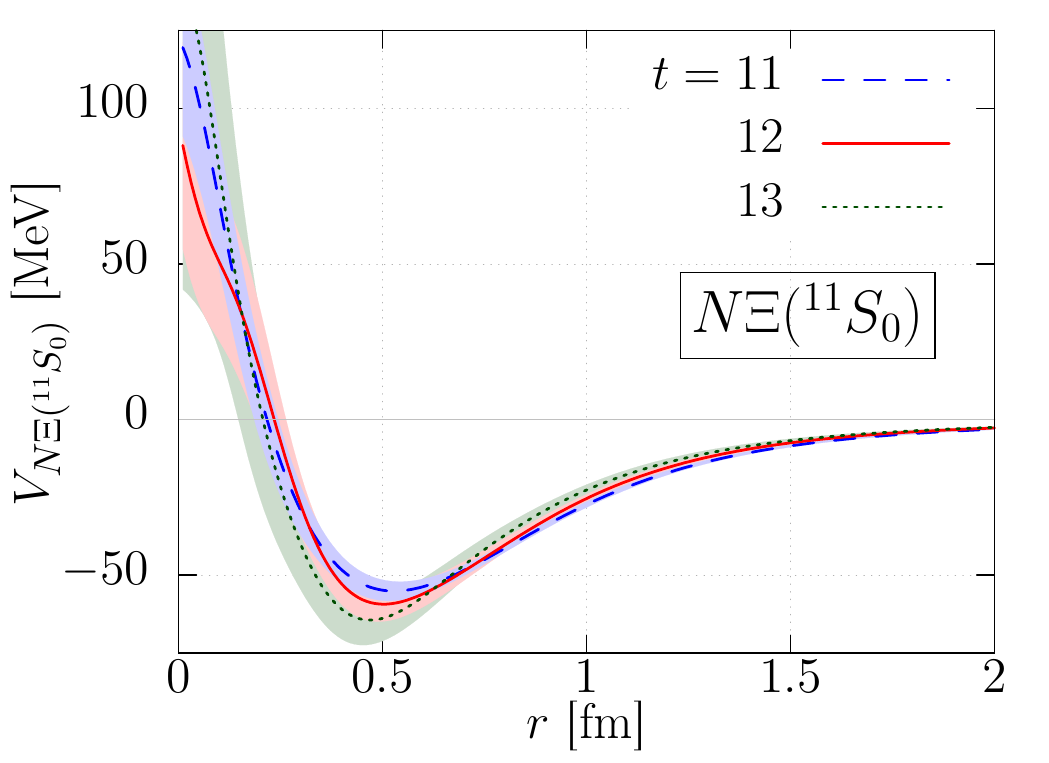}
			\includegraphics[width=0.3\textwidth]{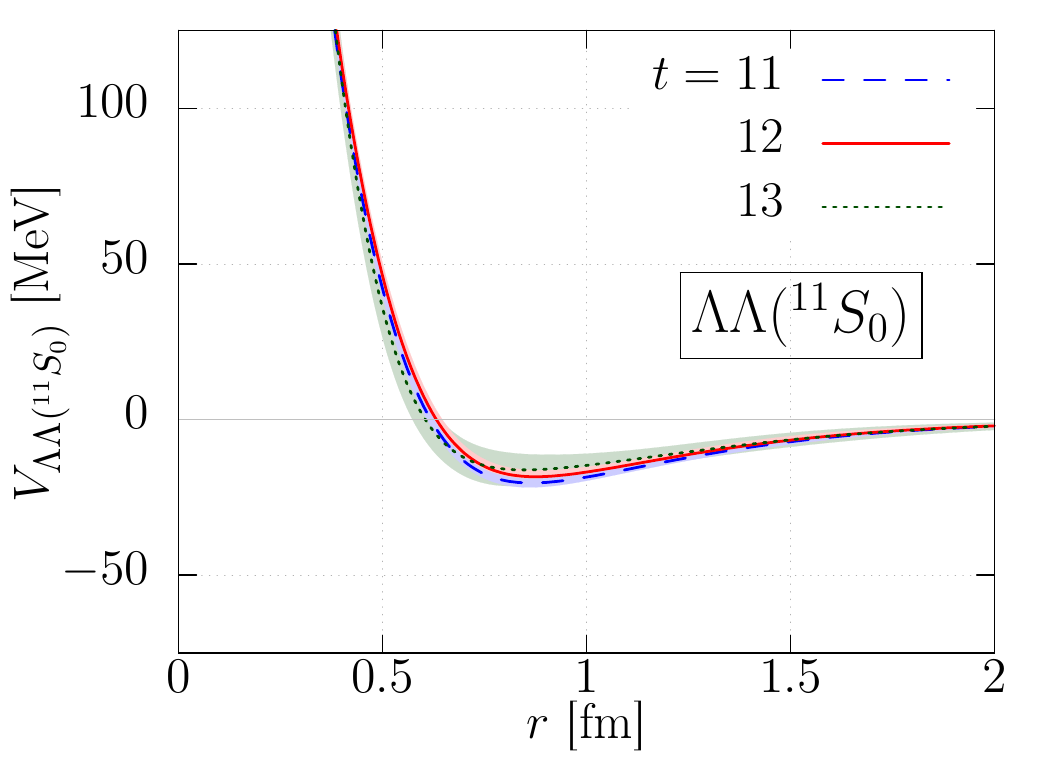}
			\includegraphics[width=0.3\textwidth]{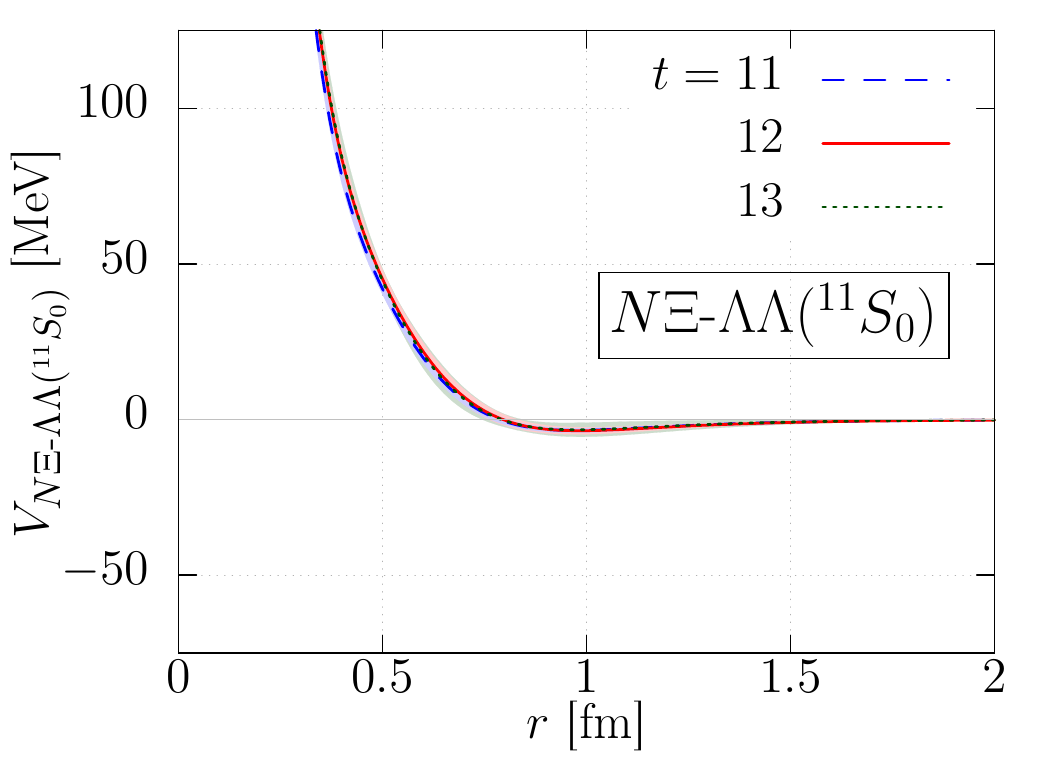}	
			\includegraphics[width=0.3\textwidth]{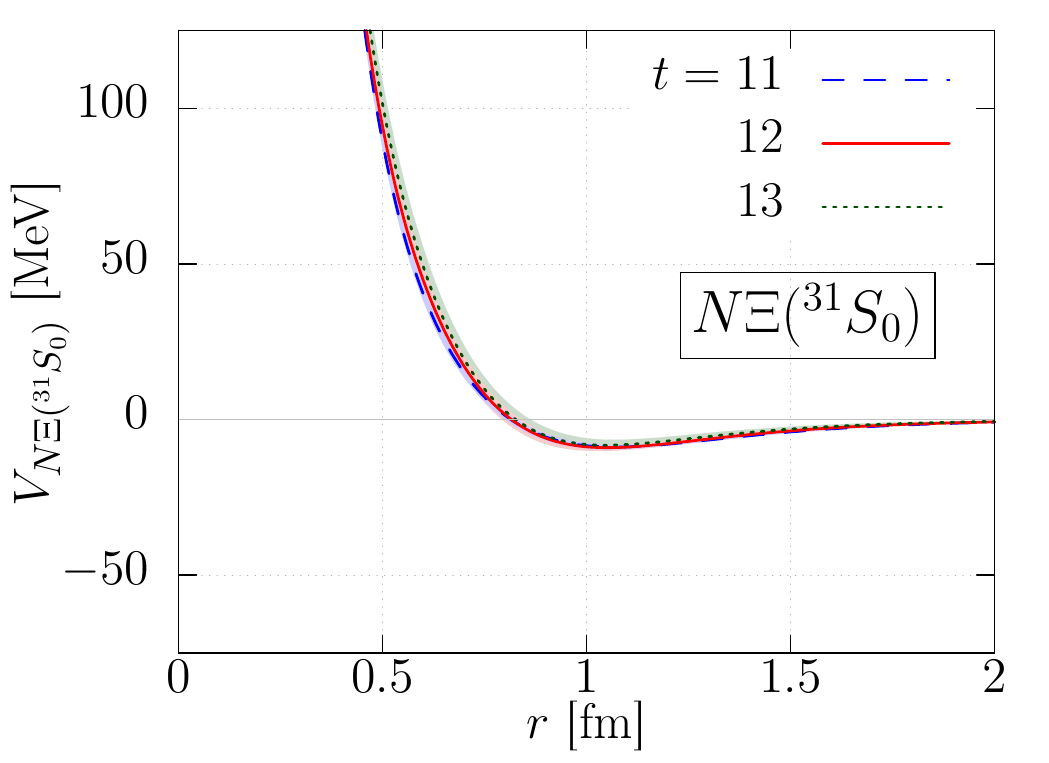}
			\includegraphics[width=0.3\textwidth]{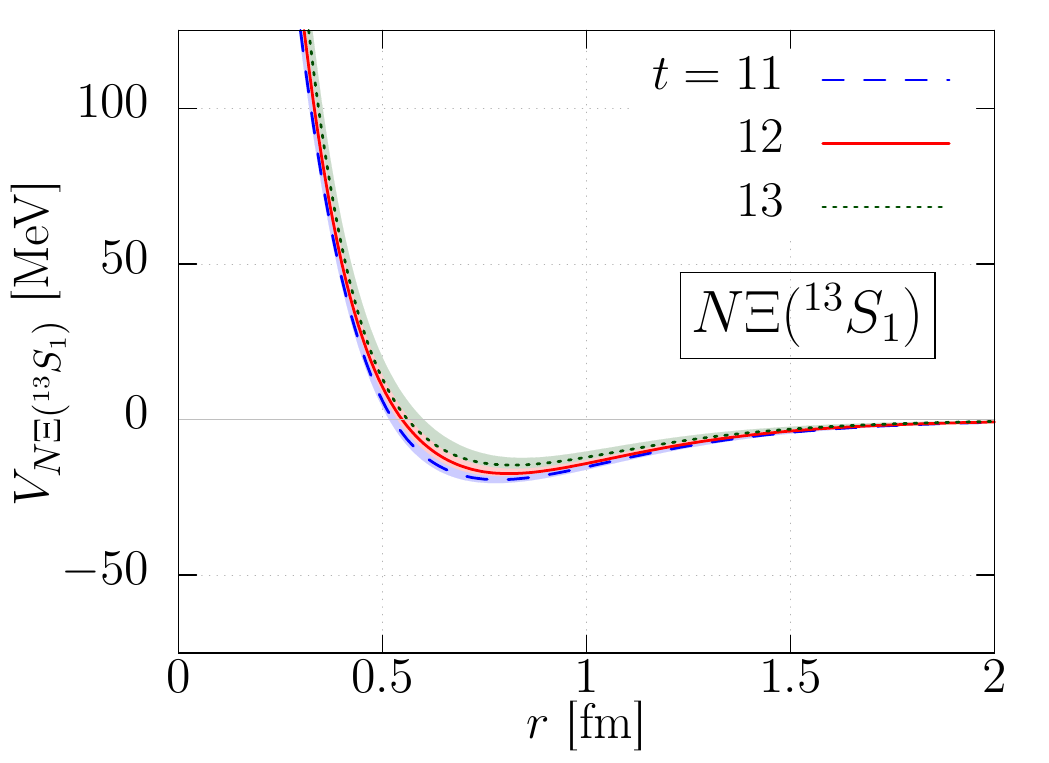}	
			\includegraphics[width=0.3\textwidth]{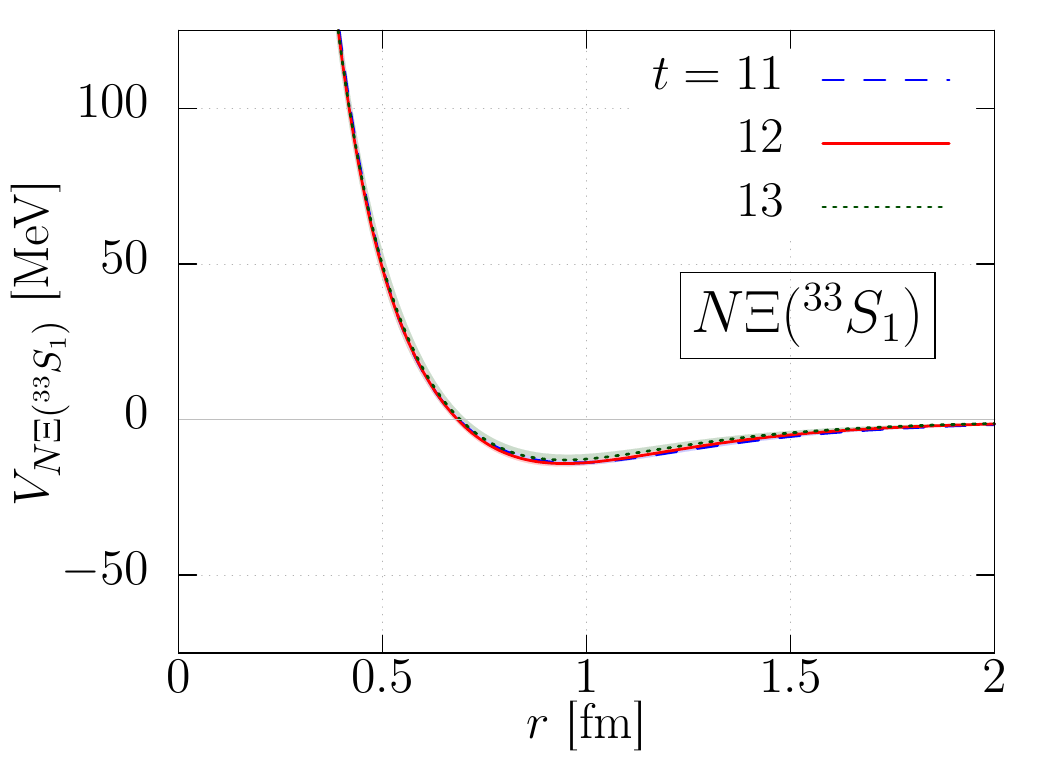}
			
		\end{minipage}
		\caption{The $s$-wave coupled-channel HAL QCD potential 
				for three temporal distances, $t=11, 12$, and 13 at almost physical quark masses \cite{Sasaki:2019qnh}.
		 The colored shadow denotes the statistical error of each potential.}
		\label{fig:HALpot}
	\end{center}
\end{figure*}

\begin{table*}
	\begin{center}
		\begin{minipage}{1\hsize}
			\begin{tabular}{|c|c|c|c|}
				\hline 
				total spin & baryon pair & $a_0$ [fm]  & $r_{\rm eff}$ [fm] \\ 
				\hline \hline
			 \multirow{3}{*}{ $J=0$ }  & $p\Xi^- $  & $-1.22 (0.13)( ^{+0.08}_{-0.00}) -i1.57(0.35)( ^{+0.18}_{-0.23})
$ & $3.7(0.3)( ^{+0.1}_{-0.1}) -i2.7(0.2)( ^{+0.1}_{-0.3}) $\\ 				
				& $n\Xi^0 $ & $-2.07(0.39)( ^{+0.28}_{-0.35}) - i0.14(0.08)( ^{+0.00}_{-0.01})$ &$1.5(0.3)( ^{+0.0}_{-0.0})-i0.2(0.0)( ^{+0.0}_{-0.1})$				\\
				& $\Lambda\Lambda$ & $-0.78(0.22)( ^{+0.00}_{-0.13})$&$5.4(0.8)( ^{+0.1}_{-0.5})$ \\ \hline 
			\multirow{2}{*}{ $J=1$ }  & $p\Xi^- $  & $-0.35(0.06)( ^{+0.09}_{-0.07})-i0.00$ & $8.3(1.0)( ^{+2.8}_{-1.2})+i0.0(0.1)( ^{+0.1}_{-0.0}) $\\ 
				& $n\Xi^0 $ & $-0.35(0.06)( ^{+0.09}_{-0.07}) 
				$ &$8.4(1.0)( ^{+2.7}_{-1.2})    $\\		 
				\hline 
			\end{tabular}
			\caption{The scattering length ($a_0$) and the effective range ($r_{\rm eff}$) 
			 in the  $p\Xi^-$,  $n\Xi^0$, and   $\Lambda\Lambda$ channels calculated by using the HAL QCD potential.
			 The Coulomb interaction is switched off. 
			  The statistical  and systematic errors are shown in the first and second parentheses, respectively.}
			\label{tab:a0_error}
		\end{minipage}
	\end{center} 
\end{table*}

Throughout this paper, we employ the state-of-the-art coupled-channel $N\Xi$-$\Lambda\Lambda$  potential below  the $\Sigma \Sigma$ threshold
 obtained by the (2+1)-flavor  lattice QCD simulations near the physical point ($m_\pi = 146\ {\rm{MeV}}$ and $m_K = 525\ {\rm{MeV}}$)~\cite{Sasaki:2019qnh}.  
 It is the local and energy-independent potential in the leading-order of the derivative expansion at low energies~\cite{Ishii:2006ec,HALQCD:2012aa}.
 The coupled-channel $N \Xi$-$\Lambda \Lambda$ potential is  fitted
  in terms of   a combination of Gaussian, Yukawa and squared-Yukawa functions 
   with the pion and kaon masses  on the lattice mentioned above~\cite{Sasaki:2019qnh}. 
        Shown in Fig.~\ref{fig:HALpot} are the results of the fitted potentials 
         in the isospin-spin basis with the notation, $^{2I+1, 2s+1} L_J$ with the isospin $I$ and the spin $s$. 
  The statistical error of the potentials originating from the Monte Carlo simulations is evaluated  by the standard jackknife method and 
    is denoted by the colored shadows, while the systematic error originating from the truncation of the derivative expansion
    is estimated  by the $t$-dependence of the potentials with $t$ being the temporal distance between 
     source and sink operators in the lattice unit  \cite{Sasaki:2019qnh}.
   The important features of the HAL QCD potential are (i) a large attraction in the $I=s=0$ $N\Xi$ channel 
 (the upper left panel), (ii)  a weak mixing between $N\Xi$ and $\Lambda \Lambda$ (the upper middle panel) at low energy, 
 and (iii) a weak  attraction in the $\Lambda \Lambda$ channel (the upper right panel).

As low energy constants characterizing the strong interaction, we calculate the scattering length $a_0$ and the effective range $r_{\rm eff}$
in the $s$-wave by solving the Schr\"odinger equation with the HAL QCD potential in Fig.\ref{fig:HALpot}  without the Coulomb interaction.
   Here we take the nuclear and atomic  physics convention, where the $s$-wave phase shift at low energies is given by
\begin{eqnarray}
q \cot \delta_0(q) = -\frac{1}{a_0} + \frac{1}{2}r_{\rm eff} q^2 + \cdots ,
\end{eqnarray}
 with $q$ being the relative momentum.
 Table \ref{tab:a0_error}   summarizes the results where  
   the central values of $a_0$ and $r_{\rm eff}$  are obtained from  $t$ = 12  with 
   the statistical errors  evaluated by the jackknife method and the systematic errors estimated from  $t = 11$ and 13.
  Unlike the procedure in Ref.~\cite{Sasaki:2019qnh} where baryon masses measured on the lattice are used in the 
  kinetic part of the the Schr\"odinger equation, we use the  experimental baryon masses of $p, n, \Lambda, \Xi^-$, and $\Xi^0$.
  \footnote{In Appendix~\ref{sec:phys_para}, we show the results of $a_0$ and $r_{\rm eff}$ with 
  the experimental baryon masses in the kinetic term and a  modified HAL QCD potential in which 
   $m_{\pi,K}$ in the fitted potential are replaced by the isospin-averaged experimental values of the pion and kaon masses.
  The results  in this procedure are consistent with those of Table \ref{tab:a0_error} within statistical and systematic errors.}

Note that $a_0$ in $\Lambda\Lambda (J=0)$ and $n\Xi^{0}(J=1)$ channels in Table \ref{tab:a0_error} are 
strictly real since there are no two-baryon states below, while those in
 $p\Xi^-(J=0)$ and $n\Xi^0(J=0)$ channels are complex due to the coupling to the lower $\Lambda\Lambda$ channel.
 Also,  $a_0$ in the  $p\Xi^-(J=1)$ channel is complex in principle due to the coupling to the lower $n\Xi^{0}(J=1)$ channel.

Solving the Schr\"odinger equation, we find that
neither bound $H$ dibaryon below the $\Lambda\Lambda$ threshold nor a 
quasi-bound state below the $N\Xi$ threshold are allowed with the HAL QCD potential, 
although  the interactions in both channels are attractive. 
Also,  the large  $|a_0|$ in the $n\Xi^0(J=0)$ channel  indicates that this system is close to the unitary regime.
In fact, there appears a virtual pole in the complex energy plane (see Appendix \ref{virtual-pole}).
The imaginary part of $a_{0}$ in the  $p\Xi^{-}$ ($J=1$) channel  is essentially zero, which implies  that the transition 
between  $p\Xi^{-}$ to $n\Xi^{0}$ is very weak: This is partly due to the fact that 
 the $N\Xi$ potential in $I=0$ (the lower middle panel  of Fig.~\ref{fig:HALpot}) and that in  $I=1$ (the lower right panel
  of  Fig.~\ref{fig:HALpot})  are very close to each other.

\section{Coupled-channel correlation function with Coulomb interaction}\label{sec:corr_model}

In high-multiplicity events of $pp$ and $p$A collisions as well as in  high-energy AA collisions, 
the hadron production yields are well described by the statistical model, which implies that 
 the hadrons are  produced independently.
 In such a situation, the momentum correlations between outgoing particles are 
generated by the the quantum statistics and the final state  interactions.
 Consider two particles, $a$ and $b$,  
with relative momentum $\bm{q} = (m_b \bm{p}_a - m_a \bm{p}_b)/ (m_a + m_b)$
observed in the final state.
Let this two-particle state be fed by a set of coupled channels,
each denoted by $j$. 
In the pair rest frame of the two measured particles,
their correlation function $C(\bm{q})$  is given by~\cite{Lednicky:1998r}
\begin{equation}
C(\bm{q})= \int d^3r \sum_j \omega_{j} S_j(\bm{r}) |\Psi^{(-)}_{j}(\bm{q};\bm{r})|^2
\ ,\label{Eq:KPLLL}
\end{equation}
where the wave function $\Psi_j^{(-)}$ in the $j$-th channel is written
as a function of the relative coordinate $\bm{r}$ in that channel,
with outgoing boundary condition on the measured channel. 
$S_j(\bm{r})$ and $\omega_j$ are the normalized source function and its weight in the $j$-th channel, respectively;
$\int d^3r S_j({\bm r}) = 1$ and $\omega_1 =1$,  where we label the measured channel as channel 1. 
The latter normalization of the source weight follows from the fact that the correlation function must be 
unity for any momentum $q$ in the non-interacting limit $V_{ij} \rightarrow 0$~\cite{Lednicky:1998r}. 
In this study, we use the static Gaussian $S_R(r)\equiv \exp(-r^2/4R^2)/(4\pi R^2)^{3/2}$ with source size $R$ for the hadron source function.
In this case, the correlation function only depends on $q=|\bm{q}|$.
Thus the correlation function contains information on both the hadron source and the hadron-hadron interactions. 
We call Eq.~\eqref{Eq:KPLLL}  the 
 Koonin{\textendash}Pratt{\textendash}Lednicky{\textendash}Lyuboshits{\textendash}Lyuboshits (KPLLL) formula 
  after the series of works~\cite{Koonin:1977fh,Lednicky:1981su,Pratt:1986cc,Anchishkin:1997tb,Lednicky:1998r}.

There are essentially four theoretical ingredients to  fully utilize the KPLLL formula and to  compare with the experimental data:
(i) the coupled channel wave functions, (ii)  threshold differences,  (iii) the modern hadron-hadron interactions, and  (iv) the Coulomb interaction. 
 A comprehensive analysis with all these ingredients   has been recently carried out
  for the first time in the case of  the $K^-p$ correlation function  in high-energy nuclear collisions
  on the basis of  the $\bar{K}N$-$\pi \Sigma$-$\pi \Lambda$ coupled-channel framework \cite{Kamiya:2019uiw}.
  In the subsections below, we generalize this approach applicable to the  $N\Xi$-$\Lambda \Lambda$ system.

\subsection{Coupled channel formalism}
 Let us first illustrate  some features of the coupled-channel wave function 
for non-identical particles.
We focus  on the small momentum region and assume
that the strong interaction modifies only  the $s$-wave part of the wave function.
The coupled-channel wave function $\Psi_{j}^{(-)}$ with the outgoing boundary condition can be written as 
\begin{equation}
\Psi^{(-)}_{j}(\bm{q};\bm{r})=(\phi({\bm q};{\bm r}) - \phi_{0}(qr)   )\delta_{1j}
+ \psi_j^{(-)}(q;r)
,\label{Eq:wf1}
\end{equation}
where $r = |\bm{r}|$,  $\phi({\bm q};{\bm r})$ is the wave function without the strong interaction,   
$ \phi_0(qr)$ is its $s$-wave component,
and $\psi_j^{(-)}(q;r)$ is the total wave function in the $s$-wave affected by the strong interaction.

The  wave function $\psi_j^{(-)}(q;r)$ in Eq.~\eqref{Eq:wf1} can be obtained by solving the coupled-channel Schr\"odinger equation,
\begin{eqnarray}
 \sum_{j} \left( -\frac{\nabla^2}{2\mu_{i}} \delta_{ij}+V_{ij}(r) \right)  \psi_{j}(q;r) = E_i \psi_{i}(q;r) ,
\label{Eq:CCSchEq}
\end{eqnarray}
where   $E_i = E - \Delta_i$ with $\mu_{i}$ and $\Delta_{i}$ representing  the reduced mass in channel $i$ and 
the threshold energy difference between channel $i$ and  channel 1, respectively.
 Since $\Delta_1=0$, we have $E=E_1$ and $q \equiv \sqrt{2 \mu_1 E} = q_1$.  
 Note that  $E_{i > 1}$ can be positive or negative depending on the  
  scattering energy, while $E \ge 0$ for physical scattering.

Unlike the case of  the standard scattering problem where 
the flux of the incoming wave is normalized,
  the outgoing wave in the measured channel needs to be  normalized in the present case 
 under  the  boundary condition;
\begin{align}
\psi_j^{(-)}(q;r) \xrightarrow[r\to\infty]  \, {1\over {2iq_j}}
\left[\delta_{1j}\, {u_{j}^{(+)}(q_jr)\over r}
+A_{j}(q) \,{u_{j}^{(-)}(q_jr)\over r}\right].
\label{eq:BC}
\end{align}
 Here $q_{j}=\sqrt{2\mu_{j}E_j }$ for open channels ($E_j \ge0$)
  and  $q_j= -i \kappa_{j}= -i \sqrt{2\mu_{j} (-E_j)}$ for closed channels ($E_j< 0$).
   Through these relations, all the momenta $q_{j}$ can be expressed as functions of $q$. 
Also, $u_{j}^{(\pm)}(q_jr)$ denotes the outgoing ($+$) or incoming ($-$) asymptotic wave;  it is  the spherical wave  $ e^{\pm iq_jr}$  
  for channels without  the Coulomb force, while the Coulomb wave function  
 needs to be used for charged particles, $u^{{\rm C}(\pm)}_j (q_jr)= \pm e^{\mp i\sigma_{j}}[iF(q_jr)\pm G(q_jr)]$
 with  $\sigma_{j}=\mathrm{arg}\Gamma(1+i\eta_{j})$, $\eta_{j}=-\mu_{j}\alpha/q_{j}$ and $F(x)$ ($G(x)$) being the regular (irregular) Coulomb wave function.

In the following subsections, we discuss  the  coupled channel treatment with 
 $p\Xi^-$ and $\Lambda\Lambda$ as measured channels.

\subsection{$p\Xi^-$ correlation function}
Let us consider the $p\Xi^-$ correlation function and assign the channel indices $i=1, 2$, and 3 
as $p\Xi^-$, $n\Xi^0$, and $\Lambda\Lambda$, respectively.
For the $p\Xi^-$ pair, there are two $s$-wave channels,  spin 0 (singlet)  and spin 1 (triplet).
The former couples to the  singlet  $n\Xi^0$ and $\Lambda\Lambda$ channels, while 
the latter couples only to  the triplet $n\Xi^0$ channel.
What we observe in experiments is the spin-averaged correlation function given by
\begin{align}
C_{p\Xi^-}(q) = \frac{1}{4}C_{p\Xi^-}^{\rm singlet}(q) + \frac{3}{4} C_{p\Xi^-}^{\rm triplet}(q).
\label{Eq:spin-sum}	
\end{align}

\begin{figure}[t]
	\begin{center}
\includegraphics[width=7 cm ]{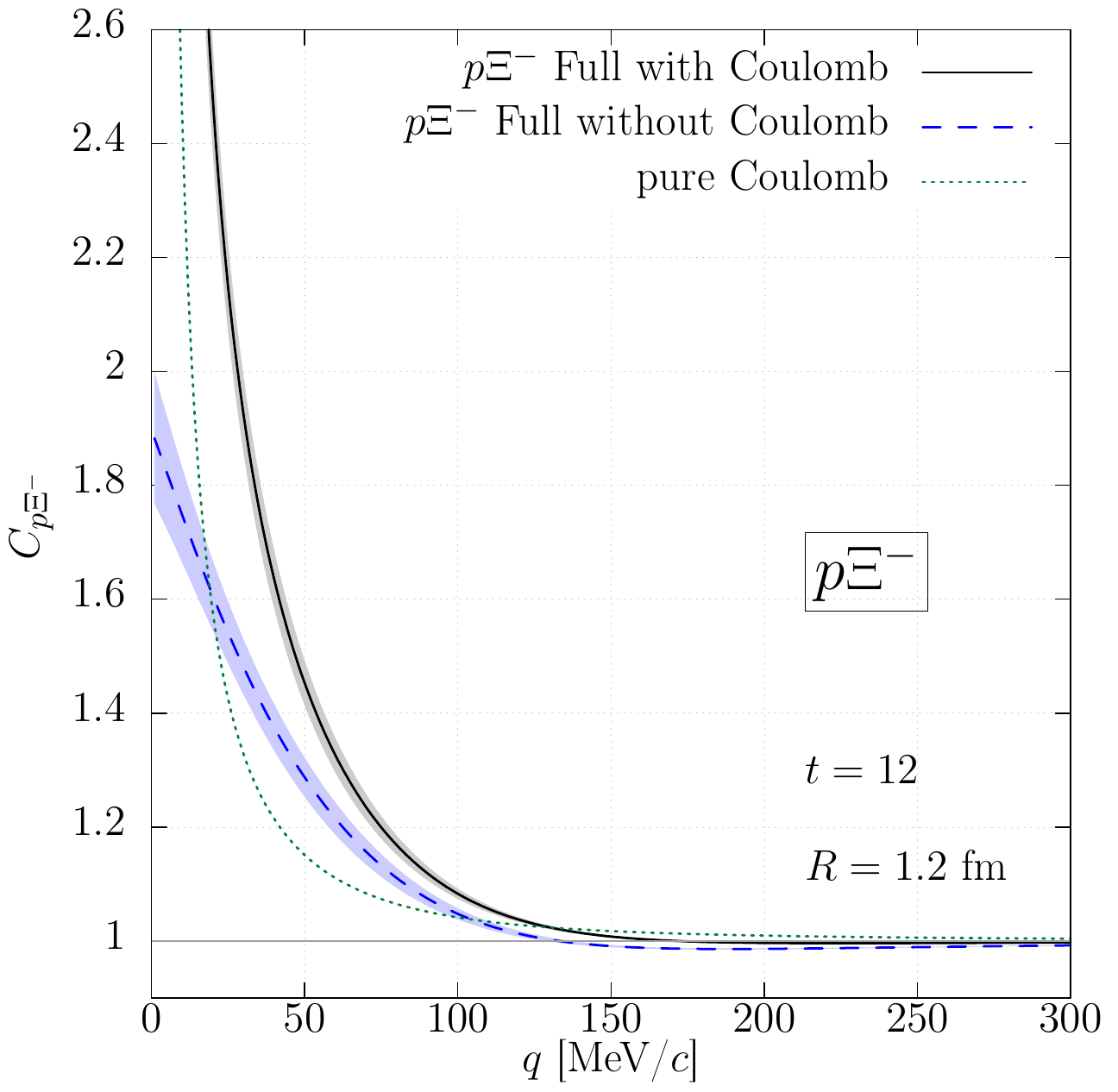}
\caption{The $p\Xi^-$ correlation function with the HAL QCD potential.  The solid (dashed) line 
 corresponds to the case with (without) the  Coulomb interaction. The statistical error from the lattice QCD data is
  shown by the shaded area. The correlation function only with the Coulomb interaction is shown by the dotted line.}
	\label{fig:pXi_JK_error}
	\end{center}
\end{figure}

\begin{figure*}[t]
	\begin{center}
		\begin{minipage}{1\hsize}
			\includegraphics[width=0.32\textwidth]{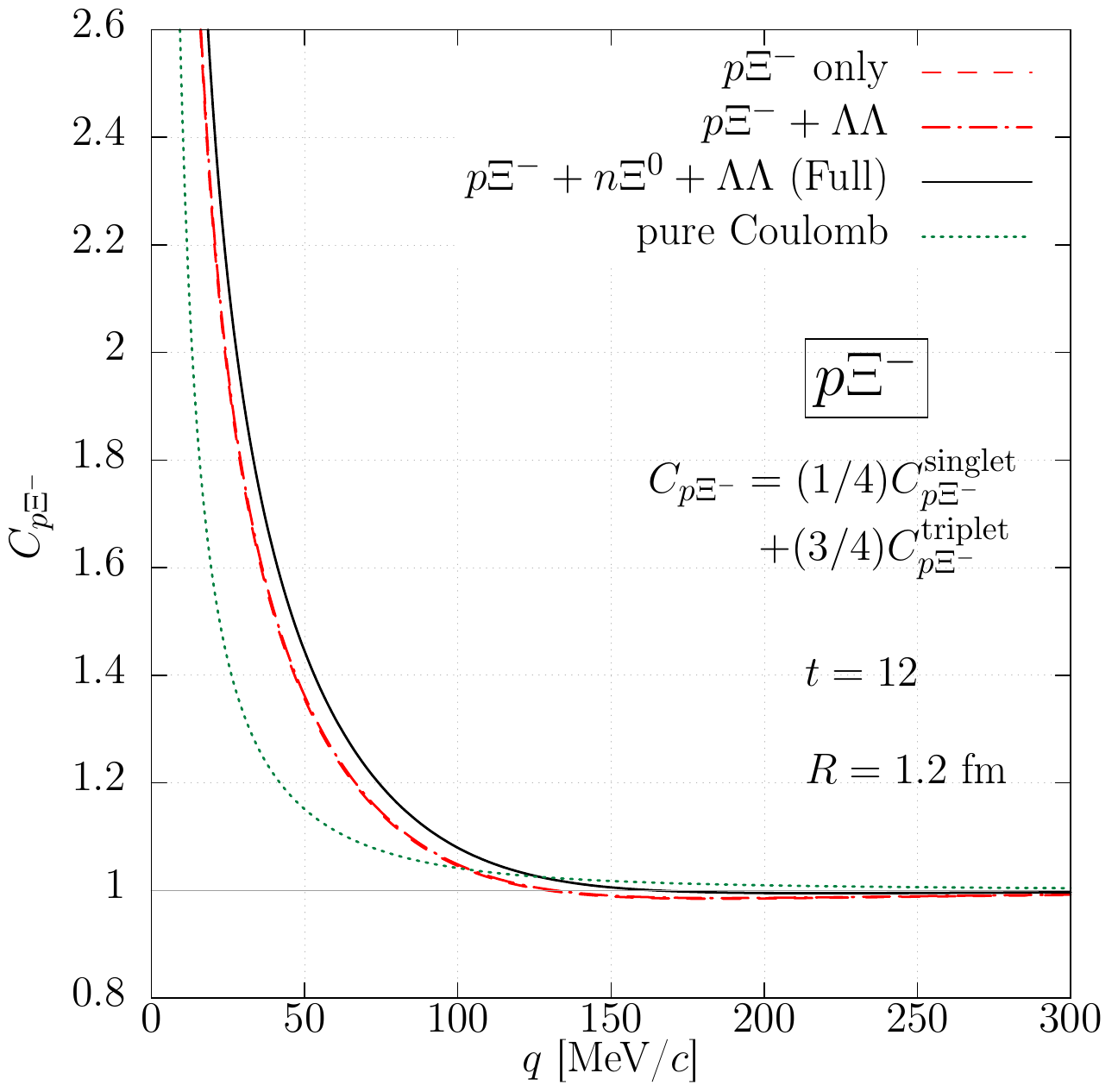}
			\includegraphics[width=0.32\textwidth]{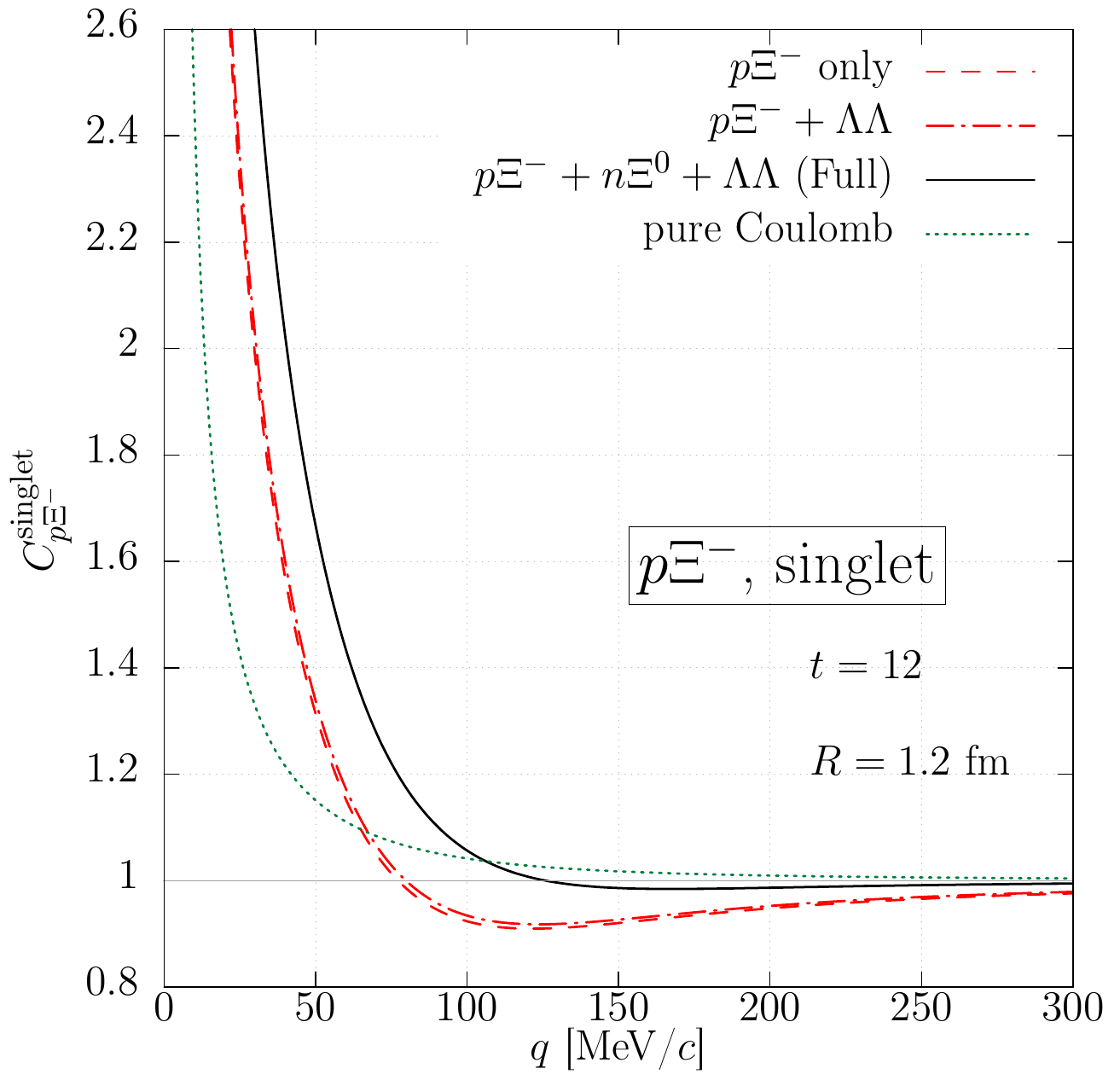}
			\includegraphics[width=0.32\textwidth]{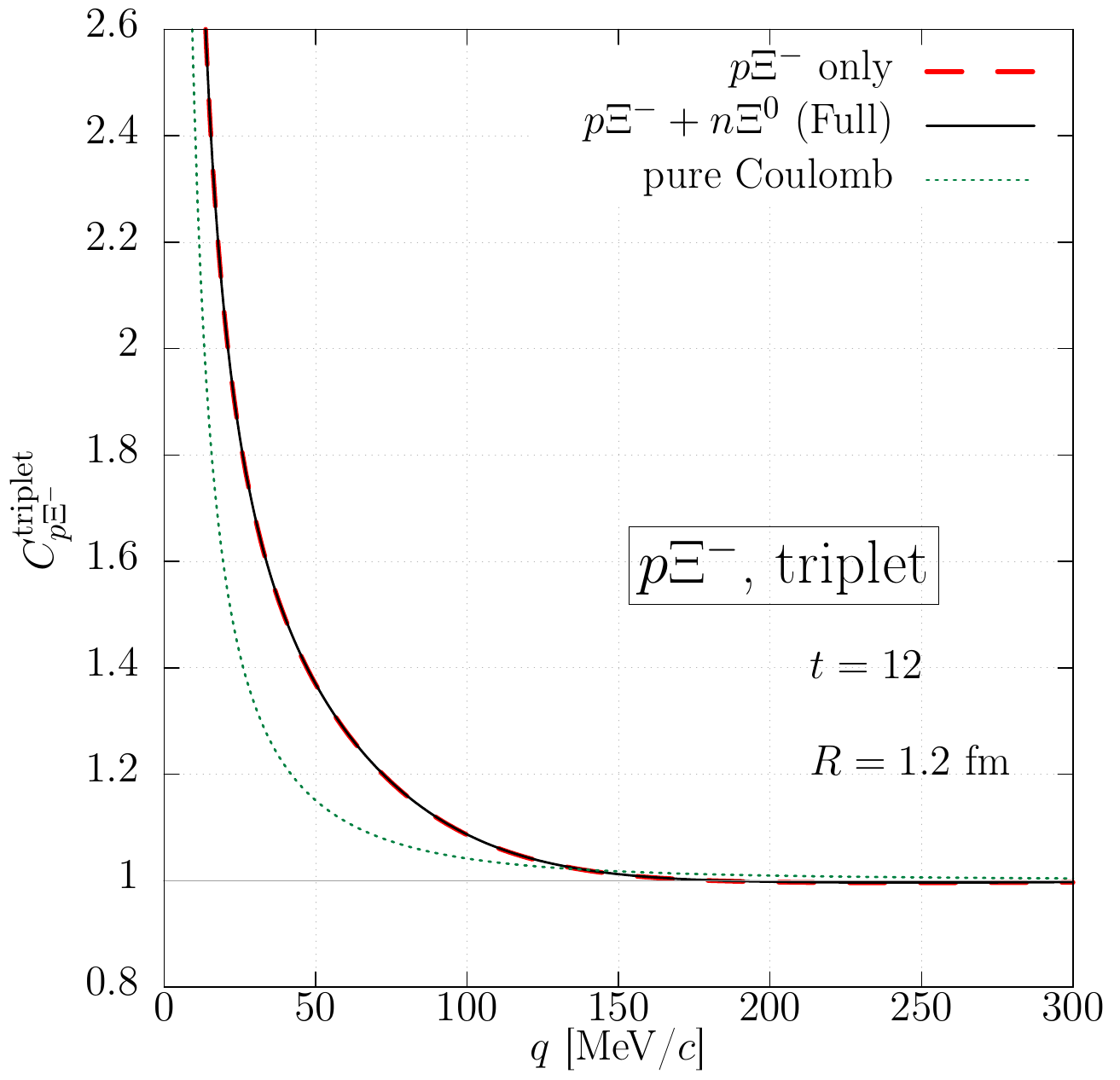}
		\end{minipage}
	\end{center}
	\caption{
The breakdown of the $p\Xi^-$ correlation function. 
The left panel shows the spin-averaged correlation function given by Eq.~\eqref{Eq:spin-sum}. 
The middle and right panels show the correlation function of 
spin single and triplet channels, respectively. 
The dashed lines denote the correlation function calculated only with the $p\Xi^-$ wave function.
The dash-dotted line and solid line denote the results
 with the contributions of $p\Xi^-\!+\!\Lambda\Lambda$ and $p\Xi^-\!+\!n\Xi^0\!+\!\Lambda\Lambda$ channels, respectively.}
	\label{fig:pXi_corr_wC}
\end{figure*}

For  $p\Xi^-$, it is necessary to treat the Coulomb interaction carefully because it distorts the wave function significantly 
in the small momentum region. We introduce the Coulomb potential $V_\mathrm{C}(r)=-\alpha/r$ to the diagonal component of the $p\Xi^-$ channel
 as $V^{\rm {QCD}}_{p\Xi^-} (r) + V_\mathrm{C}(r)$.
Since the long-range Coulomb force affects all the partial waves while the short-range strong force  affects only  the $s$-wave at low energies, the 
 wave function in  channel 1 ($p \Xi^-$)  in Eq. \eqref{Eq:wf1} should be written as ~\cite{Morita:2016auo}
\begin{align}
\Psi^{(-)}_{1}(\bm{q};\bm{r})=\left( \phi^{\rm C}(\bm{q};\bm{r})-\phi^{\rm C}_{0} (qr) \right) +\psi_1^{(-)}(q;r), \label{Eq:wf_pXi}
\end{align}
where $\phi^{{\rm C}}(\bm{q};\bm{r})$ is the free Coulomb wave function and 
$\phi^{{\rm C}}_{0}(qr)$ is its $s$-wave component.
The  boundary condition for $ \psi_j^{(-)}(q;r)$ must be given by
 the   the Coulomb wave function
   $u^{{\rm C}(\pm)}_j (q_jr)$ for  $j=1$ and the spherical wave
$ e^{\pm iq_jr}$ for $j=2$ and 3 in  Eq.~(\ref{eq:BC}).
Then the KPLLL formula  can be written as
\begin{align}
C(q)=&\int d^3r\,S_1(r)\left[|\phi^{\rm C}(\bm{q};\bm{r})|^2-|\phi^{\rm C}_{0}(qr)|^2\right]
\nonumber\\
&+\sum_{j=1}^3 \int_0^\infty 4\pi r^2dr\,\omega_{j}S_j(r)|\psi_j^{(-)}(q;r)|^2.
\label{Eq:CF_Coulomb}
\end{align}

In Fig.~\ref{fig:pXi_JK_error},
 we show the fully coupled-channel  results of the $p\Xi^-$ correlation function with and without 
the Coulomb attraction (the solid line and the dashed line, respectively), together with the case of  pure Coulomb attraction (the dotted line).
   Here we use the  $N\Xi$-$\Lambda\Lambda$ coupled-channel potential at $t=12$ given in Fig.\ref{fig:HALpot}.
 To see the qualitative behavior of $C_{p\Xi^-}(q)$, we take 
 a common source function of Gaussian shape for all channels 
   $S_j(r)=S_R(r) $  with $R=1.2$ fm and $\omega_j=1$ for all $j$. 
The error bands for the solid and dashed lines  estimated by the jackknife method reflect the statistical errors of the lattice QCD data. 
Compared to the pure Coulomb case, the correlation function shows a large enhancement by the strong interaction in the low momentum region, $q < 100$ MeV. 

To see the individual contribution in the $j$-sum in Eq. \eqref{Eq:CF_Coulomb}, 
 we   plot  in  the left panel of Fig.~\ref{fig:pXi_corr_wC}  the three cases for $C_{p\Xi^-}(q)$ with the same parameters as Fig.\ref{fig:pXi_JK_error}; 
  $j=1$ ($p\Xi^-$ only), $j=1$ and 3  ($p\Xi^- + \Lambda\Lambda$), 
 and $j=1$, 2, and 3 ($p\Xi^- + n\Xi^0 + \Lambda\Lambda$).  For simplicity, the statistical errors are not shown.
  One finds that the major enhancement of $C_{p\Xi^-}(q)$  over the pure Coulomb case comes from the $N\Xi$ attraction, 
  while the channel coupling to $\Lambda\Lambda$ is negligible.
  Further decomposition into spin-singlet $C_{p\Xi^-}^{\rm singlet}(q)$ and spin-triplet 
  $C_{p\Xi^-}^{\rm triplet}(q)$ are shown in the middle and right panels of Fig.~\ref{fig:pXi_corr_wC}, respectively.
Due to the larger negative scattering length in the spin-singlet  channel, its enhancement is stronger, although the spin degeneracy factor  is smaller.
 Also, we find that the contribution from the $n\Xi^0$ channel source to the 
singlet correlation function gives a small enhancement,
while the $\Lambda\Lambda$ source is almost negligible. 
For the triplet correlation function, the contribution from the $n\Xi^0$ source is almost invisible.

\subsection{$\Lambda\Lambda$ correlation function}
To study the  $\Lambda\Lambda$ correlation function, we  assign  the channel indices $i=1, 2$, and 3 to $\Lambda\Lambda$, $n\Xi^0$, $p\Xi^-$, respectively.
For identical particles, the wave function~\eqref{Eq:wf1} is distorted  by the quantum statistical effect. 
 Then the wave function in channel 1 ($\Lambda\Lambda$) can be decomposed in the even parity (spin-singlet) and the odd parity (spin-triplet) components as
\begin{align}
\Psi^{(-)}_{1,{\rm E}}(\bm{q};\bm{r}) &=\frac{1}{\sqrt{2}}\left[\Psi^{(-)}_{1}(\bm{q};\bm{r})+\Psi^{(-)}_{1}(\bm{q};-\bm{r})\right]\\ 
&=\sqrt{2} \left[ \left( \cos(\bm{q}\cdot \bm{r})-\phi_{0}(qr) \right)  + \psi_1^{(-)}(q;r)\right] , \label{Eq:wf_even}\\
\Psi^{(-)}_{1,{\rm O}}(\bm{q};\bm{r}) &=\frac{1}{\sqrt{2}}\left[\Psi^{(-)}_{1}(\bm{q};\bm{r})-\Psi^{(-)}_{1}(\bm{q};-\bm{r})\right]\\
& =\sqrt{2}i \sin(\bm{q}\cdot \bm{r}) .\label{Eq:wf_odd}
\end{align}
Since we consider only the $s$-wave distortion by the strong interaction, 
the scattering wave function $\psi_j$ appear only in the even parity part. 
Thus the even and odd parity correlation functions  are given by
\begin{align}
C_{\rm E}({q}) =& \int d^3r \sum_{j=1}^3 \omega_{j} S_j(r) |\Psi^{(-)}_{j,{\rm E}}(\bm{q};\bm{r})|^2\\
=&\ 1 + \exp(-4q^2R^2) \notag \\
&\! \! \! \! + 2\int d^3r \sum_{j=1}^3 \omega_{j}  S_j(r)\left[  |\psi^{(-)}_{j}(q;r)|^2-|\phi_{0}(qr)|^2\delta_{1j}\right] , \label{Eq:KP_even}\\
C_{\rm O}({q}) = &\int d^3r \sum_{j=1}^3 \omega_{j} S_j(r) |\Psi^{(-)}_{j,{\rm O}}(\bm{q};\bm{r})|^2 \\
=& 1 - \exp(-4q^2R^2).\label{Eq:KP_odd}
\end{align}

Taking into account the spin degrees of freedom, 
the final form of the $\Lambda\Lambda$ correlation function reads
\begin{align}
C_{\Lambda\Lambda}(q) =&\ \frac{1}{4}C_{\rm E}(q) + \frac{3}{4}C_{\rm O}(q)\\ 
=& \  1- \frac{1}{2} \exp(-4q^2R^2) \notag \\
&\! \! \! \! + \frac{1}{2}  \int d^3 r \sum_{j=1}^3 \omega_{j} S_j(r)\left[|\psi^{(-)}_j(q;r)|^2-|\phi_{0}(qr)|^2\right] . \label{Eq:KP_LL}
\end{align}
The $\Lambda\Lambda$ correlation is always suppressed by $(1/2) \exp(-4q^2R^2)$ by the quantum statistical effect which is 
 independent of the interactions. 

\begin{figure} 
	\begin{center}
		\includegraphics[width=0.4\textwidth]{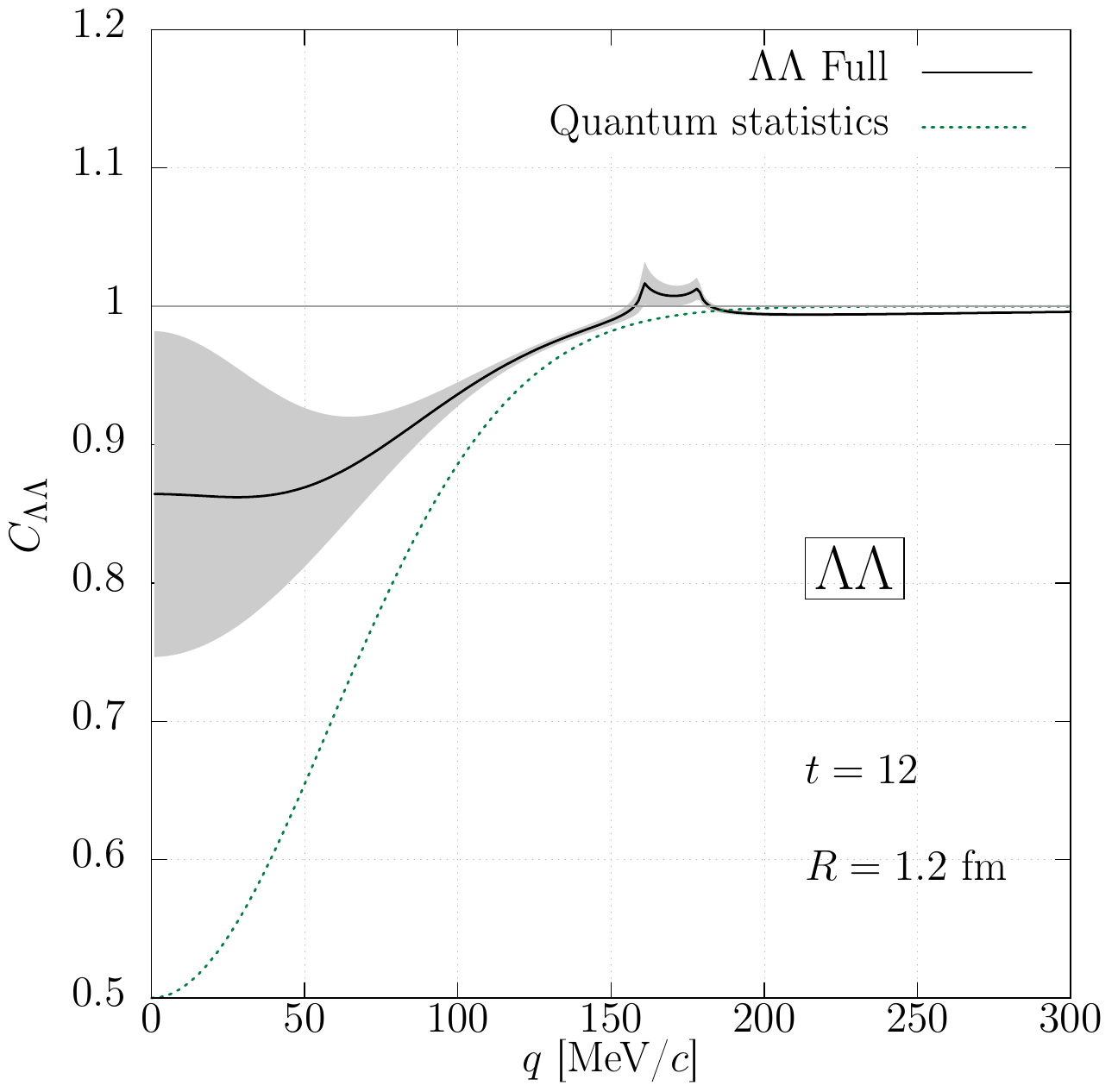}
	\end{center}
\caption{The $\Lambda\Lambda$ correlation function.
The statistical error of the lattice QCD data is shown by the shaded area.
The result of pure quantum statistics without strong interaction is shown by dotted line. 
} 
	\label{fig:LL_JK_error}
\end{figure}

We note here that, if the energy is above the $p\Xi^-$ threshold $(E_3 > 0 )$, 
$p\Xi^-$ is an open channel and the asymptotic wave function is given by the Coulomb wave function as  
\begin{eqnarray}
\psi_{3}^{(-)}(q;r)  \xrightarrow[r \rightarrow \infty]  \, {A_{3}(q)\over {2iq_{3}}}
\,{u_{3}^{{\rm C}(-)}(q_{3}r)\over r}.
\end{eqnarray}
If the energy is less than the $p\Xi^-$ threshold ($E_3 < 0$),  
 $q_3$ should be replace by $-i\kappa_3$ in the above expression, so that we have  
  $u_3^{\rm C(-)} (q_3r) = e^{i \pi |\eta_3|/2} W_{|\eta_3|, 1/2} (2\kappa_3 r)$ with
 $W_{k,\ell+1/2}(z)$ being the Whittaker function~\cite{Seaton:2002,Noble:2004eva}.

 In Fig.~\ref{fig:LL_JK_error}, we show the fully coupled-channel  result of the $\Lambda\Lambda$ correlation function (the solid line) 
 together with the case of  pure quantum statistics contribution  (the dotted line) .
 The  coupled-channel potentials at $t=12$ given in Fig.\ref{fig:HALpot} are employed, and 
a  common source function of Gaussian shape is assumed for all channels  as in the case of Fig.\ref{fig:pXi_JK_error}.
The error band for the solid line reflecting the statistical errors of the lattice QCD data is  estimated by the jackknife method. 

Compared to the case of pure quantum statistics, 
$C_{\Lambda\Lambda}(q)$  shows a strong enhancement by the strong interaction in the low momentum region: $q < 100$ MeV. 
 Also,  two cusps corresponding to the $n\Xi^0$ threshold at 2254 MeV and the $p\Xi^-$ threshold at 2260 MeV
  are found as previously pointed out in Ref.~\cite{Haidenbauer:2018jvl}.
Such a threshold cusp is indeed found experimentally in the $K^-p$ correlation 
function~\cite{Acharya:2019bsa,Haidenbauer:2018jvl,Kamiya:2019uiw}.
In the present case, these cusps are rather moderate due to the weak coupling between $\Lambda\Lambda$ and $N\Xi$, and it would be 
 a challenging problem to find them experimentally.  
 
\begin{figure}
	\begin{center}
		\includegraphics[width=7 cm]{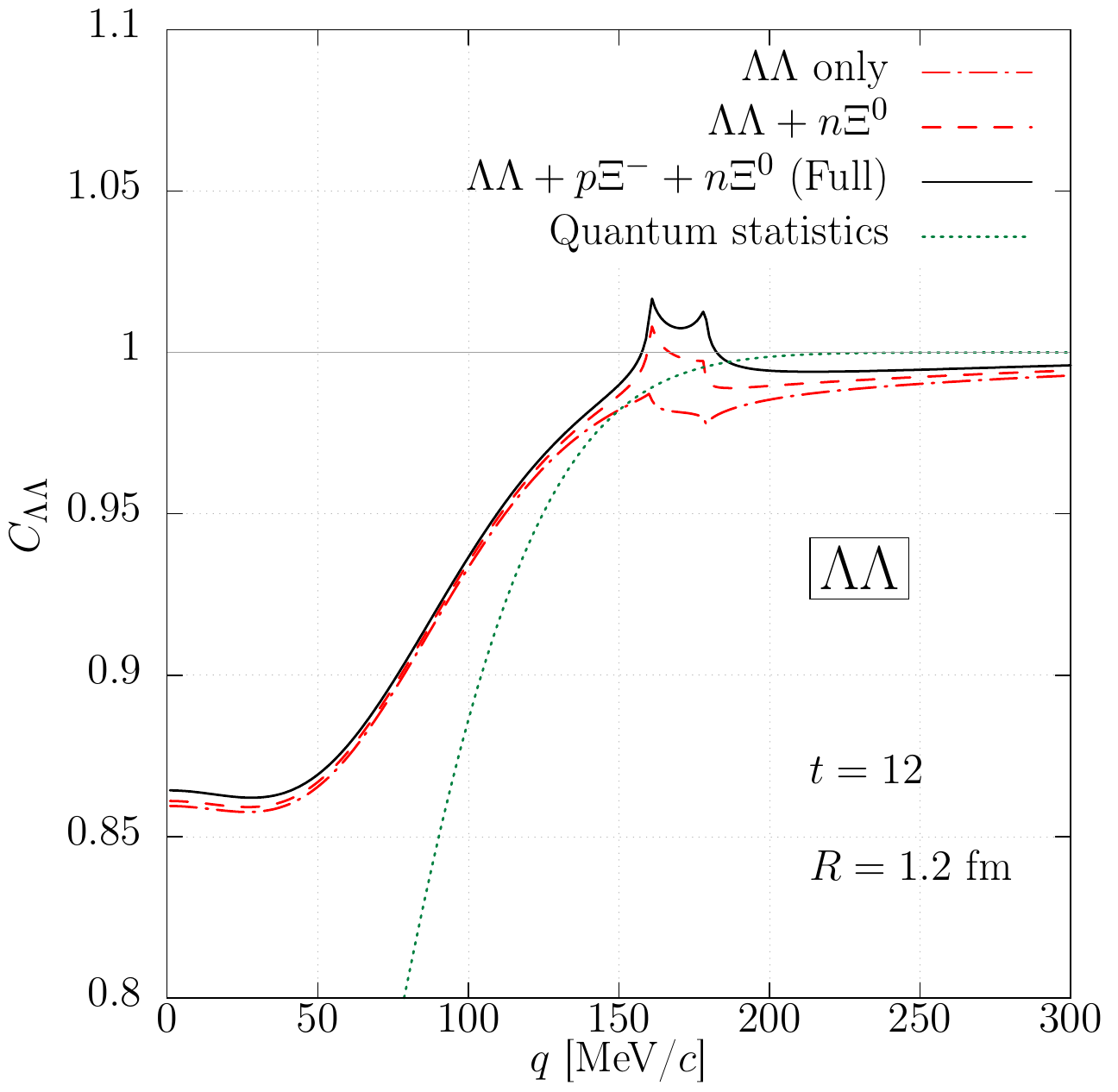}
	\end{center}
	\caption{
The breakdown of the $\Lambda\Lambda$ correlation function. 
The dashed line denotes the correlation function calculated only with the $\Lambda\Lambda$ wave function component. 
The dash-dotted (the solid line) denote the results in which the contribution from the $n\Xi^0$ (all the coupled channels) are added. 
The dotted line denotes the pure quantum statics case, where all the final state interactions are switched off.
	}
		\label{fig:LL_corr}
\end{figure}

To see the individual contribution in the $j$-sum in Eq. \eqref{Eq:KP_LL},  
 we   plot  in  the left panel of Fig.~\ref{fig:LL_corr} the three cases for $C_{\Lambda \Lambda}(q)$ with the same parameters as Fig.\ref{fig:LL_JK_error}; 
$j=1$ ($\Lambda\Lambda$ only), $j=1$ and 2  ($\Lambda\Lambda+n\Xi^0 $), 
 and $j=1$, 2, and 3 ($\Lambda\Lambda+  n\Xi^0 + p\Xi^- $).  For simplicity, the statistical errors are not shown.
The figure shows that the $n\Xi^0$ and $p\Xi^-$ sources only affect the cusp region, and make little contribution to the other momentum region. 
Nevertheless, solving the coupled-channel Schr\"odinger equation \eqref{Eq:CCSchEq} is important
 to take into  account the extra $\Lambda\Lambda$ attraction due to the coupling with $N\Xi$ states.

\section{Determination of parameters}\label{sec:comp_exp}

\subsection{Source function and weight}

For $pp$ and $p$A collisions, a spherical and static Gaussian source function works well to reproduce the data,
while the analysis of AA collisions requires more detailed information on the source function,
 e.g. asymmetrical distribution shape and flow effects~\cite{Morita:2014kza}. 
 In the present analysis of the the correlation functions in $pp$ and $p$A collisions 
 we adopt the static Gaussian source function;
\begin{eqnarray}
 S_j(r) = \frac{1}{(4\pi R_j^2)^{3/2}} \exp \left( - \frac{r^2}{4R_j^2} \right). 
\label{Eq:source-shape}
\end{eqnarray}
Here, the effective source size $R_j$ would depend on hadron pairs and reactions.
In experiments, the source size has been studied by using the
correlation function of the $pp$ pairs for which the elaborated strong interaction potential is available.
For the $pp$ pairs, the ALICE collaboration has previously determined $R_{pp}$ to be
 $R^{\text{ALICE}}_{pp}(pp)=1.182\pm0.008(\text{stat})^{+0.005}_{-0.002}(\text{syst})$ fm in  $pp$ collisions at 13 TeV  
 and $R^{\text{ALICE}}_{pp}({p\text{Pb}})=1.427\pm0.007(\text{stat})^{+0.001}_{-0.014}(\text{syst})$ fm in $p$Pb collisions
  at 5.02 TeV~\cite{Acharya:2018gyz,Acharya:2019sms}.
 On the other hand, smaller source sizes are reported for  $p\Xi^-$ and $p\Omega^-$ pairs;
 $R^{\text{ALICE}}_{p\Xi^-}(pp) = 1.02 \pm 0.05$ fm and 
  $R^{\text{ALICE}}_{p\Omega^-}(pp) = 0.95 \pm 0.06$ fm~\cite{Acharya:2020asf}.
 In the present paper, we assume that the source sizes of $N\Xi$ pairs and  $\Lambda\Lambda$ pairs are the same
 ($R_j=R$), since their total masses  are  close to each other and the contribution from the coupled channel sources is not large.

In the theoretical analysis in  Sect. \ref{sec:corr_model}, we set $\omega_j = 1$ for simplicity. 
In actual high energy collisions, the source weights depend on the channel and the reaction.
 In general, the ratio of the source weights in channels $i$ and $j$  is written in terms of the particle yields $N$ as  
 $\frac{\omega_{j}}{\omega_i} = \frac{\alpha_jN(j_1)N(j_2)}{\alpha_i N(i_1)N(i_2)}$
with $i_1, i_2, j_1 $ and $ j_2$ being the labels of particles in each channel and $\alpha_{i}$ representing the ratio of the number of particle pairs assigned to channel $i$ among $N(i_1) \times N(i_2)$ pairs.
 For the  $K^-p$ correlation function analyses in Ref.~\cite{Kamiya:2019uiw},
 a statistical  model~\cite{Borsanyi:2013bia,Bazavov:2014pvz}
  provides a reasonable estimate of the the source weights.
    Accordingly, we evaluate the ratio in terms of thermal Boltzmann factor with the corresponding baryon mass,
\begin{align}
\frac{\omega_j}{\omega_i}= \frac{\alpha_j}{\alpha_i}\exp \left( (m_{i_1} + m_{i_2} - m_{j_1} - m_{j_2})/T_{*} \right),
 \label{eq:stat_model}
\end{align}
 where  $T_* = 154$ MeV is the hadronization temperature~\cite{Borsanyi:2013bia,Bazavov:2014pvz}.\footnote{Eq.~\eqref{eq:stat_model} gives a slightly different relative weight
 	$\omega_{N\Xi}/\omega_{\Lambda\Lambda}$ from the statistical model due
 	to the approximation that holds for $m_{j_1}m_{j_2}/(m_{i_1}m_{i_2})
 	\sim 1$. We have checked that this factor does not change the
 	following qualitative results and the pictures.
 }
The factors $\alpha_{j,i}$ are given by the spin degree and the ratio of particle pairs;
$\alpha_{\Lambda\Lambda} = 1/2$ due to its identical particle nature and 
$\alpha_{N\Xi(J=0)}  = 1/4$ ($\alpha_{N\Xi(J=1)}=3/4$) due to its spin degeneracy.

\subsection{Fitting procedure}

 The experimental data of the correlation functions contain not only the physical effect of  the final state interactions
but also contaminations from the particle misidentification, the feed-down effect from weak and electromagnetic decays of other particles,
 and  the non-femtoscopic effect such as the minijet contribution.
  To take into account those effects,    we  adopt the following fitting function proposed by  the ALICE collaboration \cite{Acharya:2018gyz,Acharya:2019sms},
\begin{align}
C_{\rm fit} (q) = (a+bq) \left( 1+ \lambda(C_{\mathrm{th}}(q)-1) \right)  .
\label{Eq:Non-femt}
\end{align}
Here the first factor in the right hand side  parametrizes non-femtoscopic backgrounds. The particle misidentification and the feed-down effect are represented
 by the pair purity probability $\lambda$ which are estimated in Refs.~\cite{Acharya:2019yvb,Acharya:2019sms} and are recapitulated in
  Table~\ref{tab:non_femto_para}. 
Other correlations feeding into the present channels are assumed to be flat.
For the theoretical two-particle correlation function $C_{\mathrm{th}}(q)$, we employ the results of the HAL QCD potential  in Sect. \ref{sec:corr_model}.
We note  here that  the experimental data for the $p\Xi^-$ correlation function in $pp$ collisions  given  in Ref.~\cite{Acharya:2020asf}
are obtained after the  subtraction of the  non-femtoscopic background, the particle misidentification, and the feed-down effect,
so that we should take $(\lambda, a, b) =(1, 1, 0)$ as indicated in Table~\ref{tab:non_femto_para}.


\begin{table}
 \begin{center}
  \begin{minipage}{1\hsize}
   \begin{tabular}{|c|c|c|c|c|c|}
\hline 
collision & pair &  $\lambda$ &  $a$  & $b$ [$(\mathrm{MeV}/c)^{-1}$] & $R$  [fm]  \\ 
\hline \hline
  $pp$               & $p\Xi^-$                         &   1 ~\cite{Acharya:2020asf}          & 1 ~\cite{Acharya:2020asf}    & 0 ~\cite{Acharya:2020asf}         & \multirow{2}{*}{1.05}  \\
   (13 TeV)         & $\Lambda\Lambda$     &  0.338  \cite{Acharya:2019yvb}              &  $0.95$      & $\phantom{-}1.28\times 10^{-4}$ &                                  \\		
\hline                           		
 $p$Pb             & $p\Xi^-$                         & 0.513   \cite{Acharya:2019sms}              &   $1.09$    &  $-2.56 \times 10^{-4}$                 & \multirow{2}{*}{1.27$^{(*)}$} \\ 
   (5.02 TeV)     & $\Lambda\Lambda$     & 0.239   \cite{Acharya:2019yvb}               &   $0.99$    &  $\phantom{-}0.29 \times 10^{-4}$ &                                 \\ 
\hline	
   \end{tabular}
   \caption{The pair purity $\lambda$, non-femtoscopic parameters $a$ and $b$, and the effective source size $R$ in the 
    fitting  function $C_{\rm th}(q)$. The parameters $a$ and $b$ in $pp$ ($\Lambda\Lambda$ pairs) and $p$Pb ($p\Xi^-$ and $\Lambda\Lambda$ pairs) collisions and $R$ in $pp$ collisions are the actual fitting parameters.
    Numbers  with references are taken from Refs.~\cite{Acharya:2019yvb,Acharya:2019sms,Acharya:2020asf}, and the number with 
    $(*)$ is estimated from other 
    other parameters.  See the text for details.}
\label{tab:non_femto_para}
  \end{minipage}
 \end{center} 
\end{table}

 We carry out a simultaneous fit of  the ALICE data of $p\Xi^-$ and $\Lambda\Lambda$ correlations in $pp$ collisions
  ~\cite{Acharya:2019yvb,Acharya:2020asf} by using $C_{\rm fit}(q)$ in Eq.~(\ref{Eq:Non-femt}).
  There are three fitting parameters, $a_{\Lambda \Lambda}$, $b_{\Lambda \Lambda}$ and $R(pp)$: Other parameters are fixed 
  as given in Table~\ref{tab:non_femto_para}.  Then we found  $a_{\Lambda \Lambda}=0.95$, $b_{\Lambda\Lambda} \simeq 10^{-4} (\mathrm{MeV}/c)^{-1}$,
  and $R(pp)=1.05$ fm with $\chi^2/(\mathrm{d.o.f.}) \simeq 1$.  
 Our source size  is in  good agreement with $R^{\text{ALICE}}_{p\Xi^-}(pp)=1.02 \pm 0.05$ fm~\cite{Acharya:2020asf}. 
 
For  $p\Xi^-$ and $\Lambda\Lambda$ correlation functions in $p$A collisions ~\cite{Acharya:2019sms,Acharya:2019yvb}, 
large uncertainties of the data do not allow us to determine the source size $R({p\text{Pb}})$ precisely.  Indeed, 
  $\chi^2/(\mathrm{d.o.f.})$ depends on $R$ only weakly and stays above 1.
   Thus we estimate $R(p{\rm Pb})$  by combining our  $R(pp)$ and the ALICE results on the system dependence of the $pp$ pairs;  
$R({p\text{Pb}})=R(pp)\times R^\mathrm{ALICE}_{pp}({p\mathrm{Pb}})/R^\mathrm{ALICE}_{pp}(pp)=1.27~\mathrm{fm}$.
 After fixing $R({p\text{Pb}})$ in this way, we carry out a simultaneous fit of  the ALICE data 
 of the $p\Xi^-$ and $\Lambda\Lambda$ correlation functions in $p$Pb collisions with the four fitting parameters,
 $a_{\Lambda \Lambda}$, $b_{\Lambda \Lambda}$, $a_{p \Xi^-}$, and $b_{p\Xi^-}$ to obtain the values in  Table~\ref{tab:non_femto_para}.

Some remakes about the fitting procedure  are in order here.
(i) The statistical and systematic errors of the experimental data are added in quadrature in  our fit.
(ii) We use the data up to $q=300$ MeV/$c$ for the $p\Xi^-$ pairs in $pp$ collisions, while  
 the data up to $q=500$ MeV/$c$ are used in other cases.
  This is because  the non-femtoscopic backgrounds  are subtracted in the former case, 
  while we need high-momentum data to determine $a$ and $b$ in the latter case.
(iii) We take  the HAL QCD potential with $t=12$ to carry out the fit  $C_{\rm th}$.
 Uncertainties arising from this statistical and systematic errors of the HAL QCD potential  
 are also considered in the final results.

\section{Comparison with experimental data}\label{sec:results}

\subsection{$p\Xi^-$ correlation function}

\begin{figure*}[thbp]
 \begin{center}
  \includegraphics[width=0.5\textwidth]{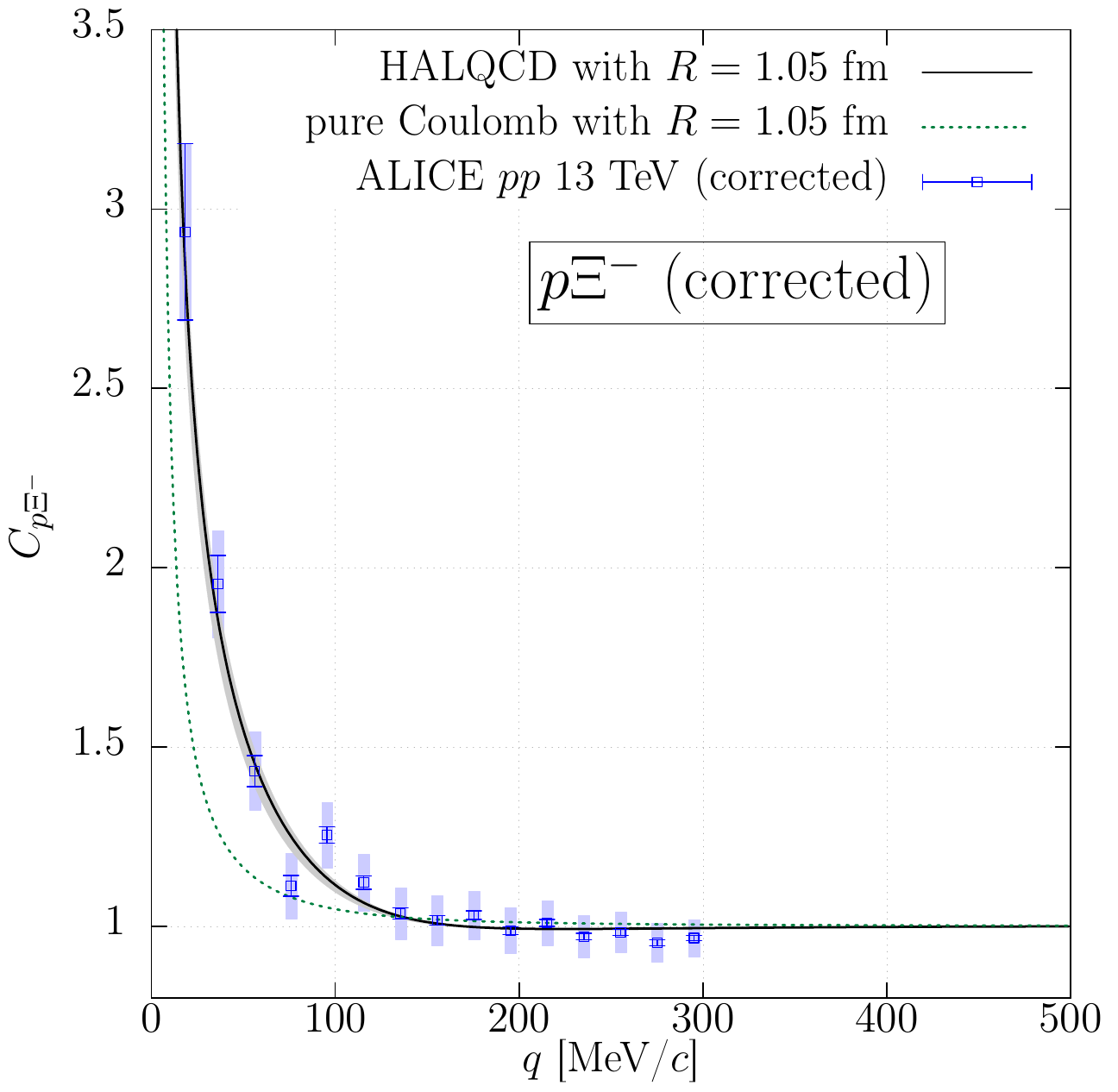}%
  \includegraphics[width=0.5\textwidth]{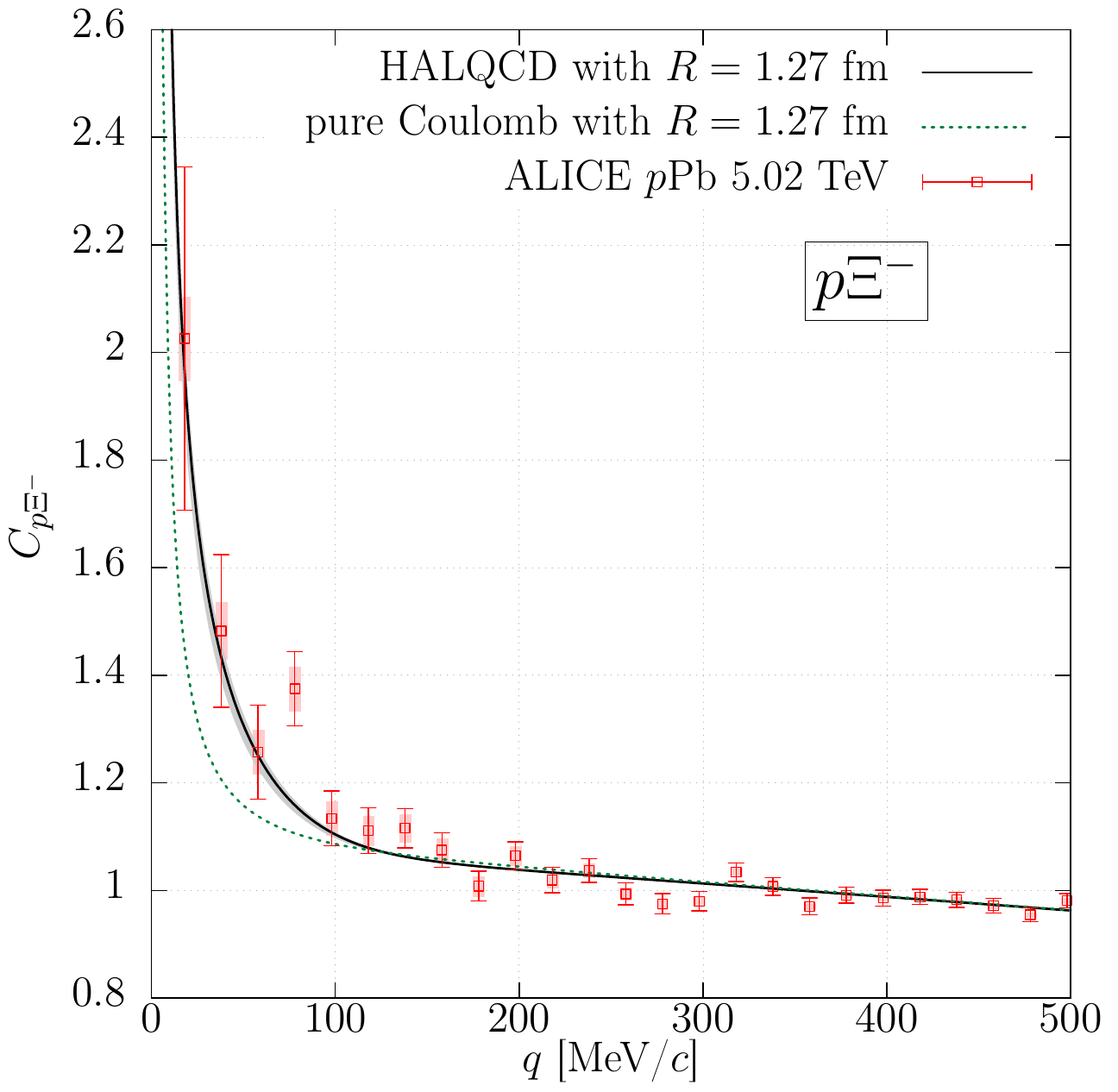}\\%
  \includegraphics[width=0.5\textwidth]{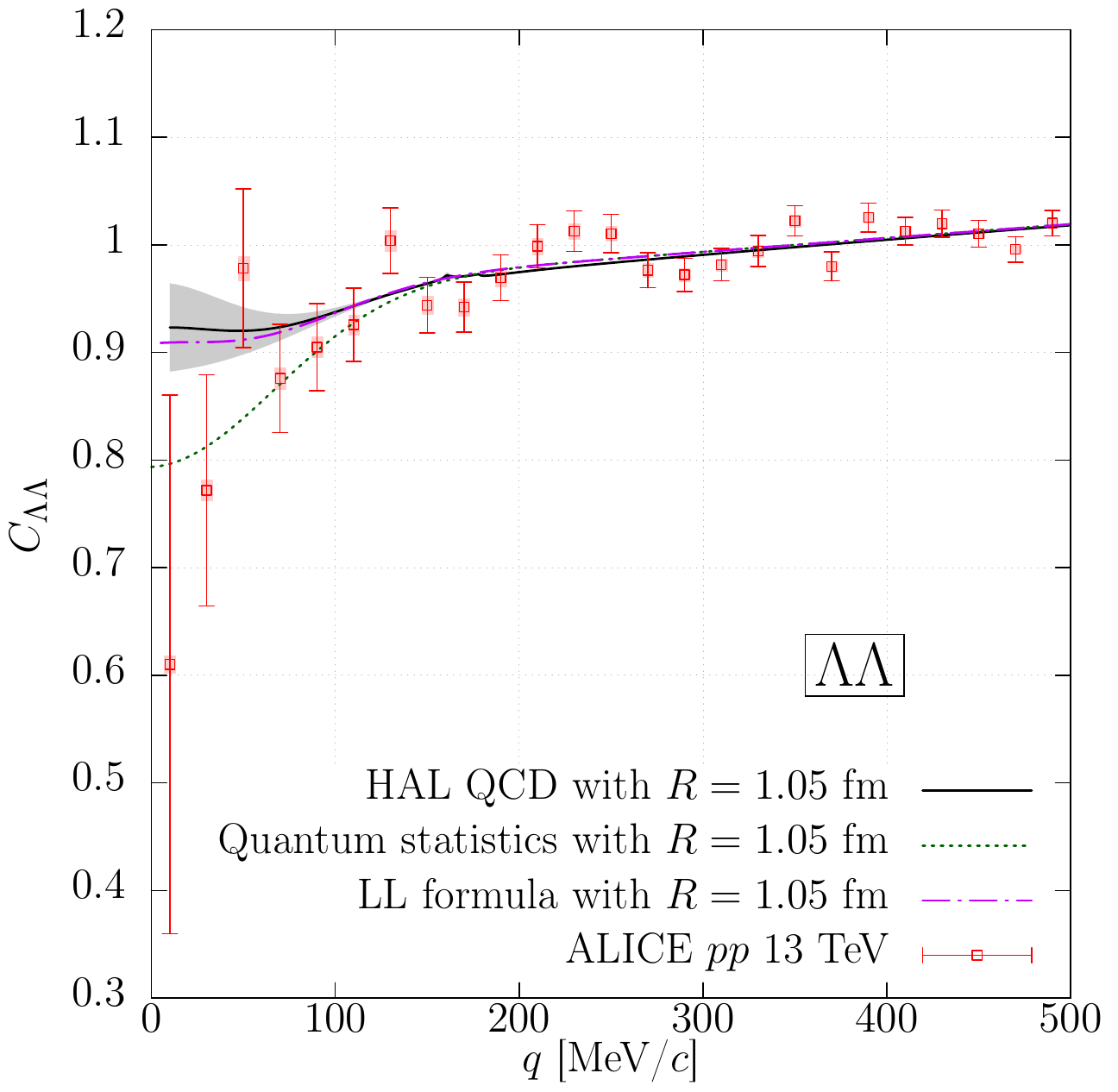}%
  \includegraphics[width=0.5\textwidth]{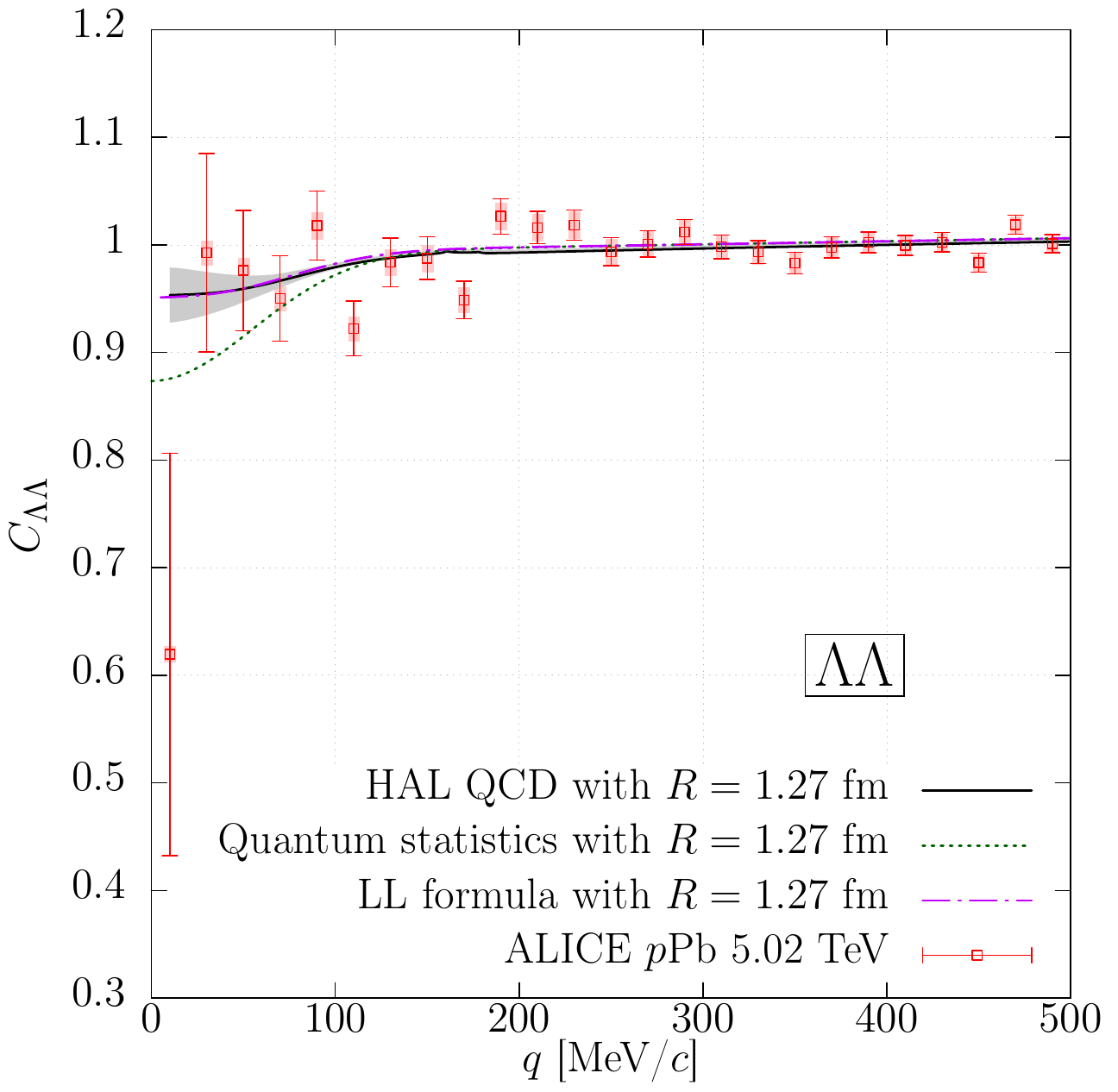}%
 \end{center}
\caption{
Experimental and theoretical correlation functions of the  $p\Xi^-$ pairs (the upper panels) and  the $\Lambda\Lambda$ pairs (the lower panels).
The blank squares are the ALICE data taken from Refs.~\cite{Acharya:2019sms,Acharya:2019yvb,Acharya:2020asf}:
 The statistical error and systematic error are denoted by the vertical line and the shaded bar, respectively.
 Solid lines are the theoretical results with with statistical and systematic  uncertainties represented by the shaded region.
 The left (right) panels correspond to the results in $pp$ collisions at 13 TeV ($p$Pb collisions ar 5.02 TeV). 
 The dotted lines show the results with only Coulomb interaction (only quantum statistics) for the $p\Xi^-$ ($\Lambda\Lambda$) correlation functions. The dash-dotted lines show the correlation function calculated with the LL formula.  
   }
\label{fig:fit_exp}
\end{figure*}

In the upper panels of Fig.~\ref{fig:fit_exp},
  our final results of the $p\Xi^-$ correlation functions are compared 
with the  $p\Xi^-$ data in $pp$ collisions at 13 TeV (the left panel) and in $p\mathrm{Pb}$
collisions at 5.02 TeV (the right panel)~\cite{Acharya:2019sms,Acharya:2020asf}. 
The solid lines denote our final results with statistical and systematic errors of the
 HAL QCD potential.  The former is estimated by the jackknife method with the $t=12$ data, 
 and the  latter is estimated by the potentials for $t=11$ and 13. The dotted green lines are the
 results with the Coulomb potential only.
Shown by the shaded region is the larger one among the statistical and systematic errors
 
The solid lines explain not only the strong enhancement at small $q$ but also 
 the $q$ dependence of $C_{p\Xi^-}(q)$.
The enhancement over  the pure Coulomb potential implies
the attractive nature of the strong $N\Xi$ interaction.
 Such an observation  has  been already reported 
  in the previous works~\cite{,Hatsuda:2017uxk,Haidenbauer:2018jvl,Acharya:2019sms,Acharya:2020asf}.
 However, our paper provides for the first time the coupled-channel analysis with the threshold difference,  the strong interaction, and the Coulomb interaction
  taken into account.  (Neither the coupled channel effect nor the threshold difference  has been   considered
   in Refs.~\cite{Hatsuda:2017uxk,Acharya:2019sms,Acharya:2020asf}, while the Coulomb interaction was not considered in Ref.~\cite{Haidenbauer:2018jvl}.)
We note that the agreement of the correlation function in Refs.~\cite{Acharya:2019sms,Acharya:2020asf}
 and that in the present work comes from the fact
 that the coupled-channel effects are not significant in the $p\Xi^-$ correlation function due to weak transition between $p\Xi^-$ and $\Lambda\Lambda$. 

\subsection{$\Lambda\Lambda$ correlation function}

In the lower panels of Fig.~\ref{fig:fit_exp},
our final results of the $\Lambda\Lambda$  correlation functions are compared 
with the  $\Lambda\Lambda$  data in $pp$ collisions at 13 TeV (the left panel) and in $p\mathrm{Pb}$
collisions at 5.02 TeV (the right panel)~\cite{Acharya:2019yvb}.
The solid lines denote our final results with statistical and systematic errors of the
 HAL QCD potential.  The dotted green lines are the   results with only the quantum statistics effect.
 Although there are large uncertainties of the experimental data at small $q$ region,
 the agreement of   the solid line with the data indicates  a weak attraction
  in the $\Lambda\Lambda$ channel without a deep  bound state.  This is
  consistent with the conclusions in Refs.~\cite{Acharya:2018gyz,Acharya:2019yvb}.

The correlation functions calculated with the Lednicky-Lyuboshits (LL) formula for identical spin-half baryon pairs~\cite{Lednicky:1981su} are also plotted 
in the lower panels of Fig.~\ref{fig:fit_exp} by the dash-dotted line: 
\begin{align}
	C (q)=&1- \frac12\,e^{-4q^2R^2} + \frac{1}{2}\Delta C(q),\\ 
\Delta C(q) = &
        \frac{|f(q)|^2}{2R^2}F_3\left(\frac{r_\mathrm{eff}}{R}\right)
	+\frac{2\mathrm{Re} f(q)}{\sqrt{\pi}R}F_1(2qR) \nonumber\\
	& -\frac{\mathrm{Im} f(q)}{R} F_2(2qR), \label{Eq:LL}
\end{align}
where $F_1(x)=\int_0^x dt\,e^{t^2-x^2}/x$, $F_2(x)=(1-e^{-x^2})/x$, $F_3(x)=1-x/2\sqrt{\pi}$, and 
 we make the effective range expansion of single channel $\Lambda\Lambda$ scattering amplitude  $f(q)$  
 with $a_0 =-0.78$ fm and $r_{\rm eff} = 5.4$ fm given in Table~\ref{tab:a0_error}.
The same non-femtoscopic parameters and the pair purity listed in Table.~\ref{tab:non_femto_para} are used.
  We find that the single-channel LL formula gives a good approximation to the fully coupled-channel results for wide range of $q$
  in both $pp$ and $p$Pb collisions.
 It would be interesting  to see whether high precision data for $C_{\Lambda \Lambda} (q)$ in the future
  may reveal cusp structures  at the $n \Xi^0$ and $p\Xi^-$ thresholds  as expected from  the coupled channel effect.

\subsection{System size dependence}
\label{sec:discussion}

\begin{figure}[tbhp]
\begin{center}
\includegraphics[width=0.48\textwidth]{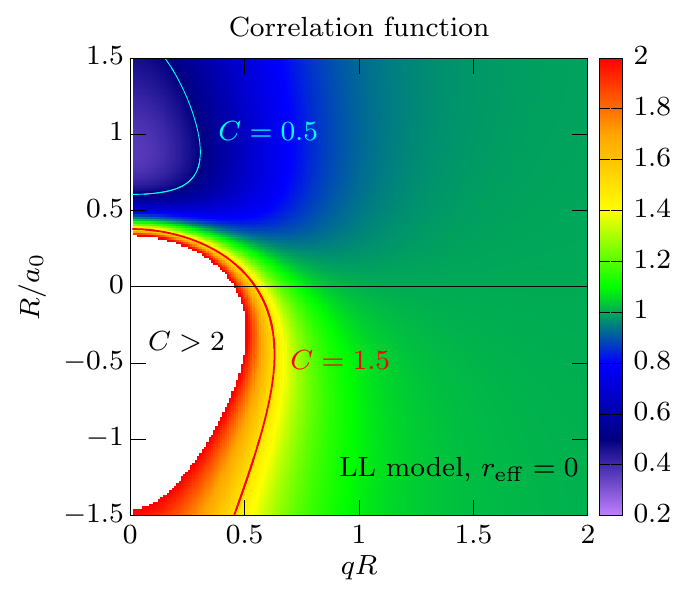}
\end{center}
\caption{The contour plot of the correlation function $C(q)$  in the LL analytic model at $r_\mathrm{eff}=0$
as a function of $x=qR$ and $y=R/a_0$.}
\label{fig:LLcorr}
\end{figure}

The enhancement of $C(q)$ for fixed  $R$ alone 
cannot conclude  whether bound or quasi-bound state is generated by the strong interaction.
  This can be demonstrated by using an analytic  model for neutral and non-identical particles $C(q) = 1+ \Delta C(q)$
  with  $r_{\rm eff}=0$  which is obtained from Eq.~(\ref{Eq:LL}) as 
   \begin{align}  
\Delta C(q)=\frac{1}{x^2+y^2}\left[\frac12-\frac{2y}{\sqrt{\pi}}F_1(2x)-x F_2(2x)\right] ,
\end{align}
with  $x=qR$ and $y=R/a_0$.
 Shown in Fig.~\ref{fig:LLcorr} is a contour plot of $C(q)$ in the $x$-$y$ plane.
The strongly enhanced region $C(q)>2$ indicated by the white area
  extends to both negative and positive sides of  $y$ for $x < 0.5$.
   (Even if one introduces the  Coulomb attraction  such as the case of $p\Xi^-$,
this situation does not change qualitatively  as discussed in Appendix C.)

\begin{figure}[tbhp]
\begin{center}
\includegraphics[width=7 cm]{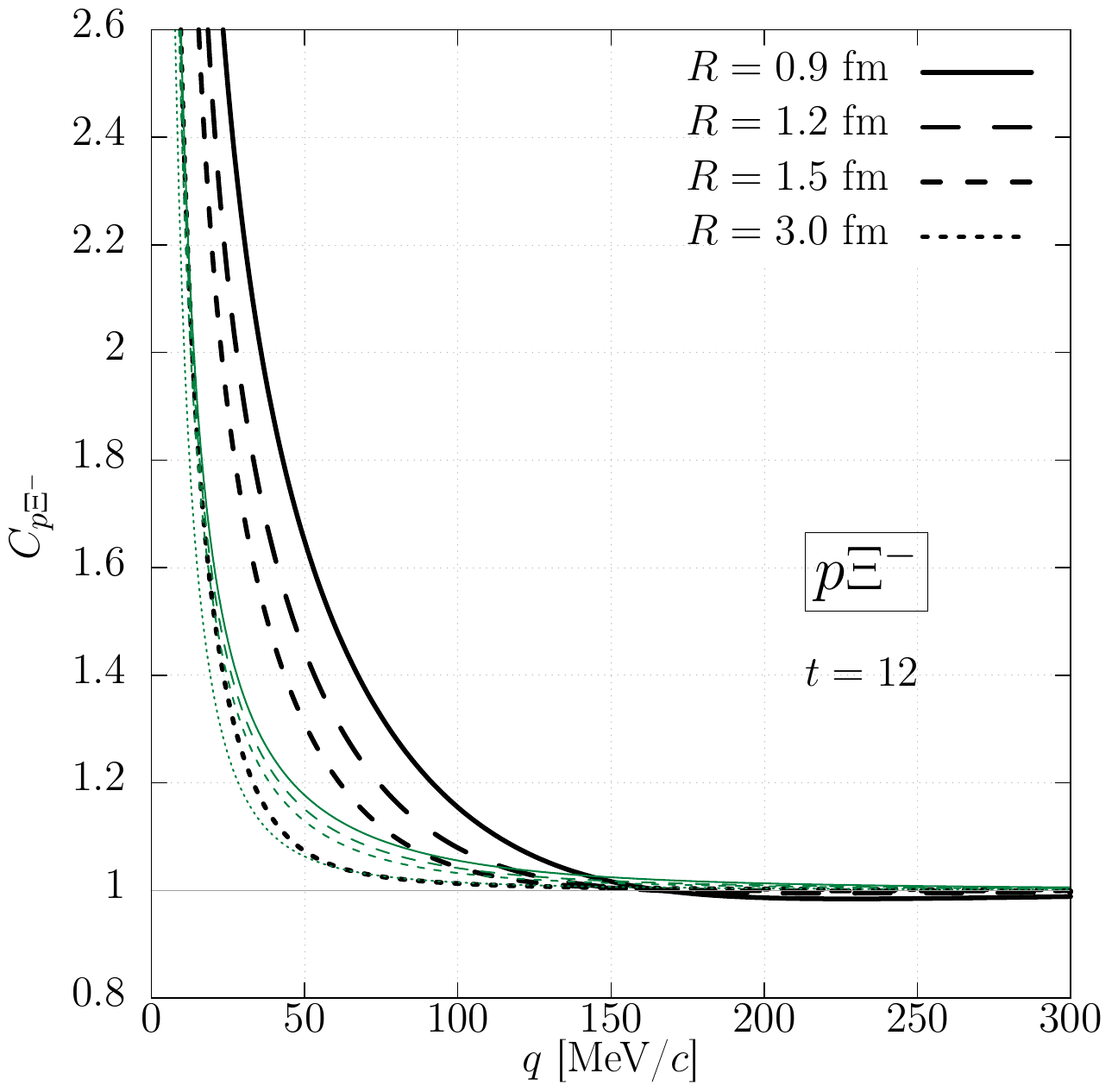}
\includegraphics[width=7 cm]{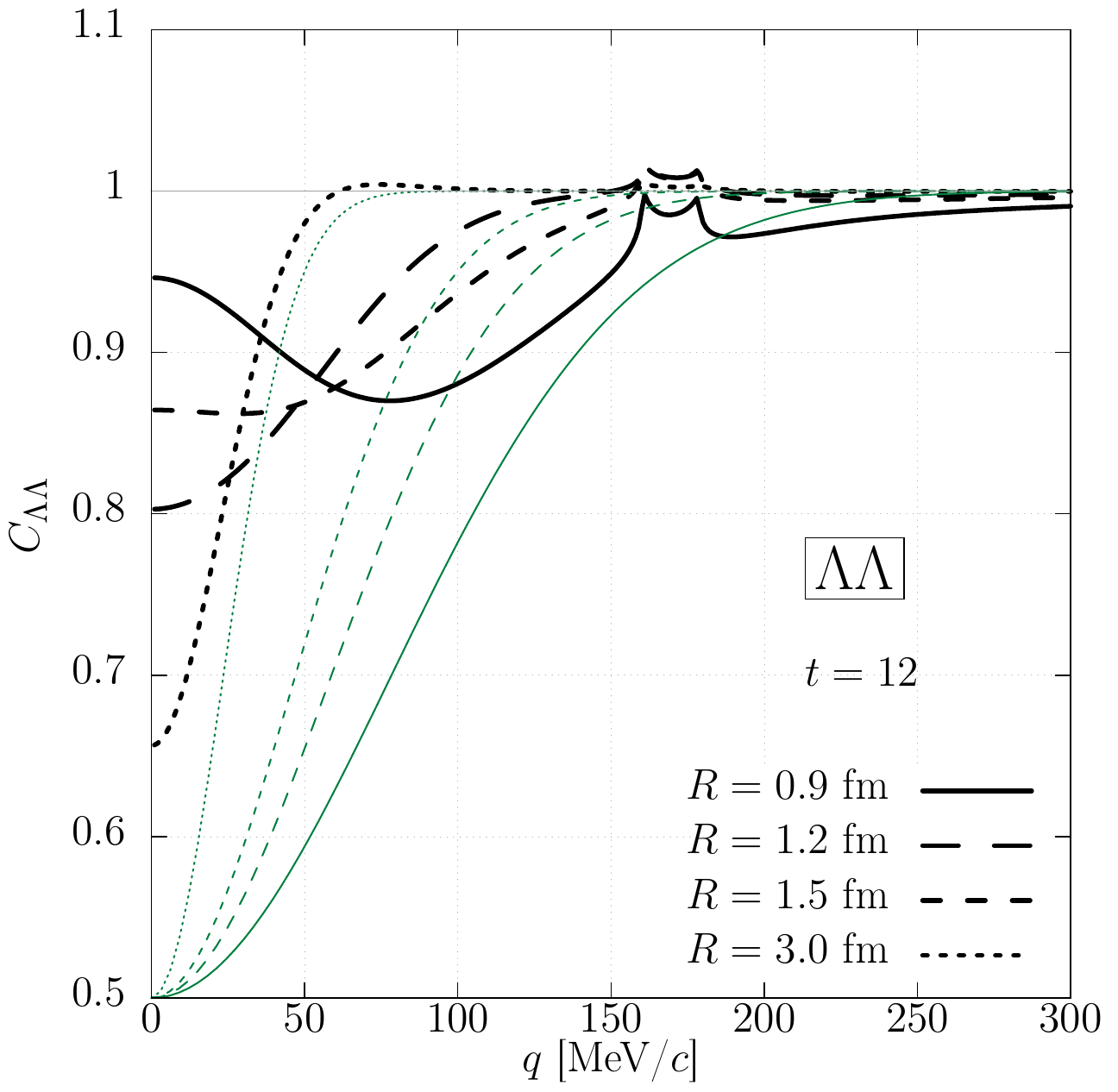}
\end{center}
\caption{Source size dependence of the $p\Xi^-$ and $\Lambda\Lambda$ correlation functions. The thick lines denote the results with full coupled-channel calculation. For comparison, the calculations with the pure Coulomb cases (pure quantum statics cases) are shown for $p\Xi^-$ ($\Lambda\Lambda$) correlation function by thin lines. 
}
\label{fig:Rdep}
\end{figure}

Scanning through the $y$-axis by changing the system size $R$ 
 would provide further  experimental information  on the sign of $y$.  
 To demonstrate this, we show the $p\Xi^-$ and $\Lambda\Lambda$ correlation
functions for several different  source sizes ($R=0.9, 1.2, 1.5,$, and$\ 3~\mathrm{fm}$) in Fig.~\ref{fig:Rdep}
 with the HAL QCD potential (the thick lines) and without the HAL QCD potential (the thin lines).
 
For  the $p\Xi^-$ correlation function, Fig.~\ref{fig:Rdep} implies that
 the enhancement of $C(q)$ due to  strong interaction
 over the pure Coulomb attraction 
 is significant around $R=1$ fm but is gradually reduced toward the larger values of $R$.
  This is consistent with the  fact that we are in the negative $y$ region as indicated by  Fig.~\ref{fig:LLcorr}.
  If the scattering length is in the bound region ($y=R/a_0 > 0$),  we would expect that $C(q)$ undershoots the
   Coulomb contribution and may form a dip as a function of $x=qR$. 
  Thus the experimental studies of the $p\Xi^-$ correlation function in heavy-ion collisions corresponding to larger $R$
   are of  particular interest.

For  the $\Lambda\Lambda$ correlation function,
Fig.~\ref{fig:Rdep} shows that  the enhancement of $C(q)$ due to strong interaction
 over the pure quantum statistics has characteristic non-monotonic behavior for $q$ smaller than the 
  $N\Xi$ threshold.  However, to make quantitative discussions for large $R$ corresponding
  to the heavy-ion collisions,  more realistic source shape as well as the flow effect 
   need to be taken into account~\cite{Morita:2014kza},   since the effect of quantum statistics 
   is particularly important in the  $\Lambda\Lambda$ correlation.

We note here that a high-momentum tail of the $\Lambda\Lambda$ correlation function
 above the $N\Xi$ threshold was observed in Au+Au collisions at RHIC~\cite{Adamczyk:2014vca},
  and a residual source having a small size ($R_{\rm res} \simeq 0.5~\mathrm{fm}$) was introduced in
   previous works~\cite{Morita:2014kza,Adamczyk:2014vca,Ohnishi:2016elb}.
 Although it was suggested in Ref.~\cite{Ohnishi:2021ger} that 
 the coupled-channel effects may explain the high-momentum tail in Au+Au collisions,  
  the present analysis shows that such a  tail does not appear  unless $R$ is smaller than  1 fm  as shown in Fig.~\ref{fig:Rdep}.
 Thus  this issue is still left open for future studies.

\section{Summary}\label{sec:summary}
We studied the $p\Xi^-$ and $\Lambda\Lambda$ femtoscopy  in $pp$ and $p$Pb collisions at LHC
  by using the latest $N\Xi$-$\Lambda\Lambda$ coupled-channel HAL QCD  potential. 
  A moderate $N\Xi$ attraction of this potential produces a virtual pole below the $n\Xi^0$ threshold. 
 On the basis of the KPLLL formula for the momentum correlations of hadron pairs, 
 we considered the coupled-channel effect, the threshold difference, the strong interaction, and the
Coulomb interaction at the same time  to analyze the $p\Xi^-$ and $\Lambda\Lambda$  correlation functions. 
After evaluating  the parameters of the non-femtoscopic effects and the source function,
theoretical results of the correlation functions are compared with the  experimental data by the ALICE collaboration;
 they are found to be in good quantitative agreement. 
From this comparison,  we concluded that the negative  scattering lengths in the $N\Xi$ system 
is implied by the strong enhancement of the $p\Xi^-$ correlation function over the Coulomb contribution.
Also, we  found that the $\Lambda\Lambda$ correlation function may show a  twin cusp near the $n\Xi^0$ and $p\Xi^-$ thresholds due to
  channel coupling,  which would be interesting to be seen in future  high precision data.
   
Studies with  femtoscopic techniques in different collision systems
 will help us to unravel the physics of hadron-hadron interactions further.
 For example, it is interesting to examine the $N\Xi$ correlation function in nucleus-nucleus collisions by changing the impact parameter, so that
 one can utilize the idea of the  ``small-to-large ratio"  to extract the strong interaction effect without much contamination from the Coulomb interaction ~\cite{Morita:2016auo}. A femtoscopic study of the hadron-deuteron correlation functions ~\cite{Mrowczynski:2019yrr,Etminan:2019pru,Haidenbauer:2020uew,Ogata:2021juh}
is another feasible and valuable direction to pursue.
 The production of the $S=-2$ system through the ($K^-, K^+)$ reaction with nuclear target is also an alternative and 
  promising approach to study the $N\Xi$-$\Lambda\Lambda$ system in a controlled
   fashion \cite{Ahn:1998fj,Yoon:2007aq,Ahn:2017}.

\noindent
{\it Acknowledgements:} 
The authors thank members of HAL QCD Collaboration
for providing lattice QCD results of coupled-channel
$N\Xi$-$\Lambda \Lambda$ interactions and for valuable discussions.
This work has been supported in part by the Grants-in-Aid for Scientific Research
from JSPS (Grant numbers
JP21H00121, 
JP21H00125, 
JP19H05150, 
JP19H05151, 
JP19H01898, 
JP19K03879, 
JP18H05236, 
JP18H05236, 
JP18H05407, 
JP16H03995, 
and
JP16K17694), 
by the Yukawa International Program for Quark-hadron Sciences (YIPQS),
by a priority issue (Elucidation of the fundamental laws and evolution of the universe)
to be tackled by using Post ``K" Computer,
by  ``Program for Promoting Researches on the Supercomputer Fugaku'' (Simulation for basic science: from fundamental laws of particles to creation of nuclei),
by the Joint Institute for Computational Fundamental Science (JICFuS),  
by the National Natural Science Foundation of China (NSFC) under Grant No.~11835015 and No.~12047503, 
by the NSFC and the Deutsche Forschungsgemeinschaft (DFG, German Research Foundation) through the funds provided to the Sino-German Collaborative Research Center TRR110 {\lq\lq}Symmetries and the Emergence of Structure in QCD'' (NSFC Grant No.~12070131001, DFG Project-ID 196253076), 
by the Chinese Academy of Sciences (CAS) under Grant No.~XDB34030000 and No.~QYZDB-SSW-SYS013, the CAS President's International Fellowship Initiative (PIFI) under Grant No.~2020PM0020, and China Postdoctoral Science Foundation under Grant No.~2020M680687.

\appendix

\section{Low energy constants from modified HAL QCD potential}\label{sec:phys_para}

\begin{table*}[t]
	\begin{center}
		\begin{minipage}{1\hsize}
			\begin{tabular}{|c|c|c|c|}
				\hline 
				total spin & baryon pair & $a_0$ [fm]  & $r_{\rm eff}$ [fm] \\ 
				\hline \hline
 \multirow{3}{*}{ $J=0$ }  & $p\Xi^- $                  & $  -1.25(0.03)( ^{+0.12} _{-0.00})-i2.00(0.40)( ^{+0.16} _{-0.31})$  & $3.7(0.3)( ^{+0.0} _{-0.1})-i2.4(0.2)( ^{+0.1} _{-0.3})$\\ 				
           & $n\Xi^0 $                 & $-2.76(0.63)( ^{+0.33} _{-0.66})-i 0.15(0.12)( ^{+0.00} _{-0.03})$ &$ 1.5(0.3)( ^{+0.0} _{-0.1}) -i0.1(0.0)( ^{+0.0} _{-0.0})$\\
           & $\Lambda\Lambda$ & $-0.99(0.30)( ^{+0.00} _{-0.17})$ &$4.9(0.70)( ^{+0.1} _{-0.5})$ \\ \hline 
 \multirow{2}{*}{ $J=1$ }  & $p\Xi^- $                  & $ -0.47(0.08)( ^{+0.11} _{-0.09})-i0.0(0.00)( ^{+0.00} _{-0.00})$         &$ 6.7(0.7)( ^{+1.4} _{-0.9})+i 0.0(0.1)( ^{+0.0} _{-0.0})$\\ 
  		   & $n\Xi^0 $                 & $ -0.47(0.08)( ^{+0.11}_{-0.09})	$ &$ 6.8(0.7)( ^{+1.4}_{-0.9})$\\
				\hline 
			\end{tabular}
			\caption{Same with Table \ref{tab:a0_error} but with the pion and kaon masses in the 
			fitted HAL QCD potential  in Ref.~\cite{Sasaki:2019qnh} by
			the  isospin averaged physical masses,  $m_{\pi}=137.3$ MeV and
			$m_K=495.6$ MeV.}
			\label{tab:a0_error_phys_m}
		\end{minipage}
	\end{center} 
\end{table*}

The HAL QCD potential used in the text  is constructed at  $m_\pi \simeq 146 \physdim{MeV}$ and $m_K \simeq 525\physdim{MeV}$ 
which are slightly away from the physical point ~\cite{Sasaki:2019qnh}.
To estimate the effect of this discrepancy, we replace $m_{\pi}$ and $m_K$  in the parametrization of the HAL QCD potential by
 the isospin-averaged physical masses, 137.3 MeV and 495.6 MeV, respectively.
  Resulting scattering lengths and effective ranges are shown in Table~\ref{tab:a0_error_phys_m}.
  The numbers are consistent with those given in Table~\ref{tab:a0_error} within the errors, although the central values of the scattering length $a_0$ are
   slightly larger due to slight increase of the attraction by the smaller pion and kaon masses.

\section{Virtual pole and $N\Xi$ interaction}\label{virtual-pole}

\begin{figure*}[t]
\begin{center}
\begin{minipage}{1\hsize}
	\centering{\includegraphics[width=0.75\textwidth]{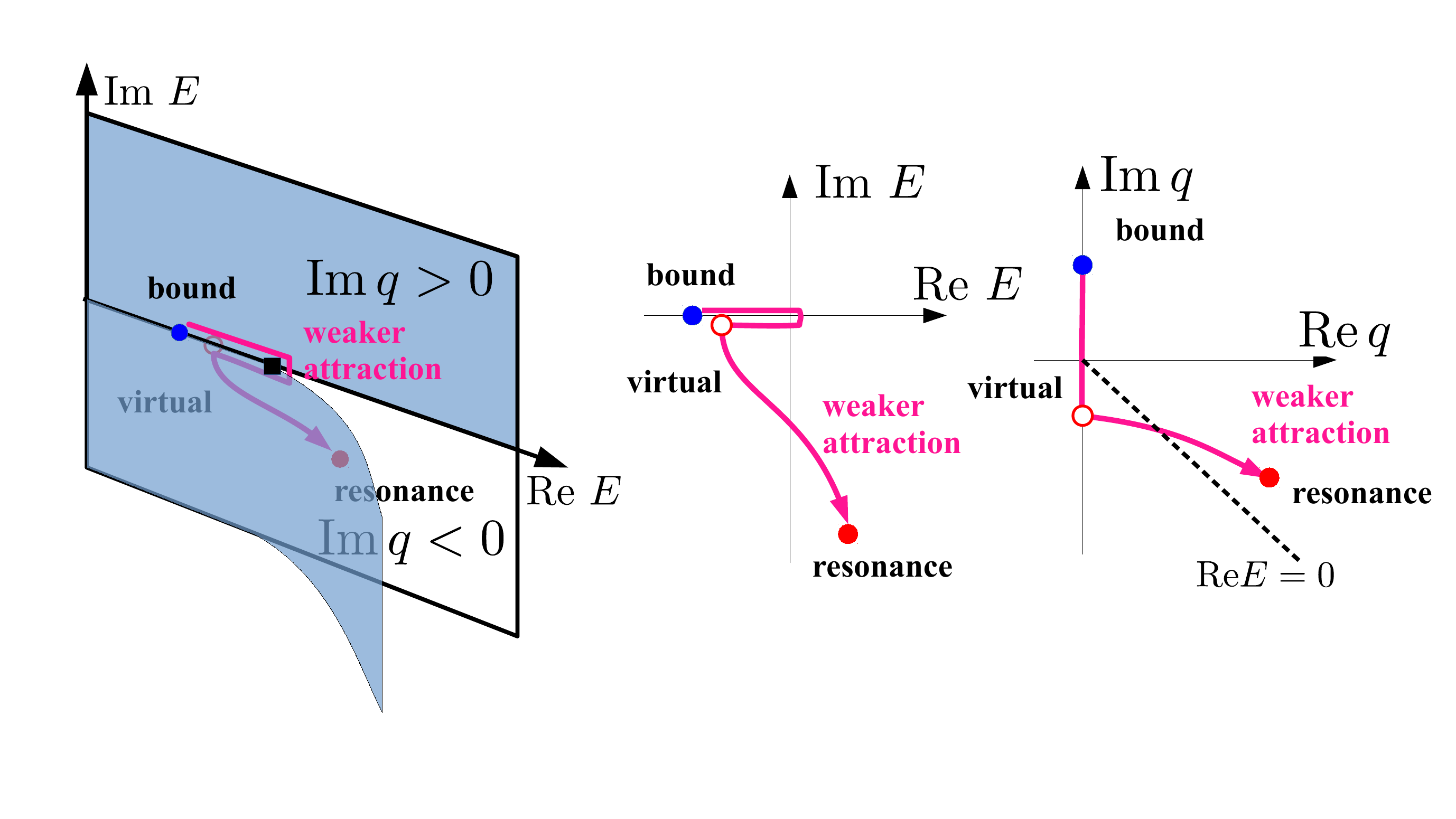}}
\end{minipage}
	\caption{A schematic picture of the $s$-wave pole position generated by the strong interaction
		in the complex energy and momentum space.  As the attractive interaction becomes weaker from the bound region,
		 the bound pole becomes the virtual pole  first, and then moves on to the resonance pole. }
	\label{fig:virtual}
\end{center}
\end{figure*}

In the single-channel problem with a sufficiently attractive $s$-wave  interaction,
a bound state pole lies on the imaginary momentum $q$ axis
in the upper half of the complex $q$ plane, as shown in Fig.~\ref{fig:virtual}.
This pole goes down along the imaginary $q$ axis to the lower half plane (virtual pole),
and eventually goes off the imaginary $q$ axis to the right half plane 
with decreasing attraction~\cite{Masui:2000mug}. 
 Eventually, the physical eigenstate emerges as a resonance 
  in the case where the eigen-momentum of pole $q_{\mathrm{pole}}$ 
  satisfies the condition, $\mathrm{Re}  \ q_{\mathrm{pole}} > -\mathrm{Im} \ q_{\mathrm{pole}}$.
When there are decay channels, the relation between the interaction and the pole position is 
more complicated compared to the single-channel case.
Nevertheless, we can find a similar behavior for a pole lying nearby the threshold energy.

As indicated by the negative scattering length in Table I in the case of HAL QCD potential with  $t= 12$,
the strong interaction does  not generate  bound or quasi-bound states near the  $\Lambda\Lambda$, $n\Xi^0$ and $p\Xi^-$ thresholds.
Instead, we find a virtual pole lying at $E_{\mathrm{pole}}=2250.5 - i0.3$ MeV 
in the $(+,-,+)$ sheet  in the $J=0$ channel:
The real part of the energy  is just below the $n\Xi^0$ threshold by $-3.93$ MeV, while
the sign of the imaginary part of the eigen-momenta of $\Lambda\Lambda$, $n\Xi^0$, and $p\Xi^-$, are $+$, $-$, and $+$, respectively.
If the $N\Xi$ quasi-bound state would emerge, the corresponding pole should have appeared 
 below the $n\Xi^0$ threshold in the $(-,+,+)$ sheet. 
 The near-threshold virtual pole in the $(+,-,+)$ sheet still contributes to the enhancement of  the scattering length
   in the  $n\Xi^0$ channel.
  We note  here that,   if we use the modified HAL QCD potential associated with Table~\ref{tab:a0_error_phys_m},
   the  virtual pole moves closer to the threshold energy of $n\Xi^0$, $E_{\mathrm{pole}} = 2251.8 - i0.2$ MeV.  
  This is due to the fact that the attraction becomes  slightly stronger in this case and the virtual pole moves toward right direction in the 
   complex $E$ plane as seen from Fig.~\ref{fig:virtual}.

\section{LL model with Gamow factor}\label{sec:gamow}

\begin{figure}[t]
\begin{center}
	\centering{\includegraphics[width=0.5\textwidth]{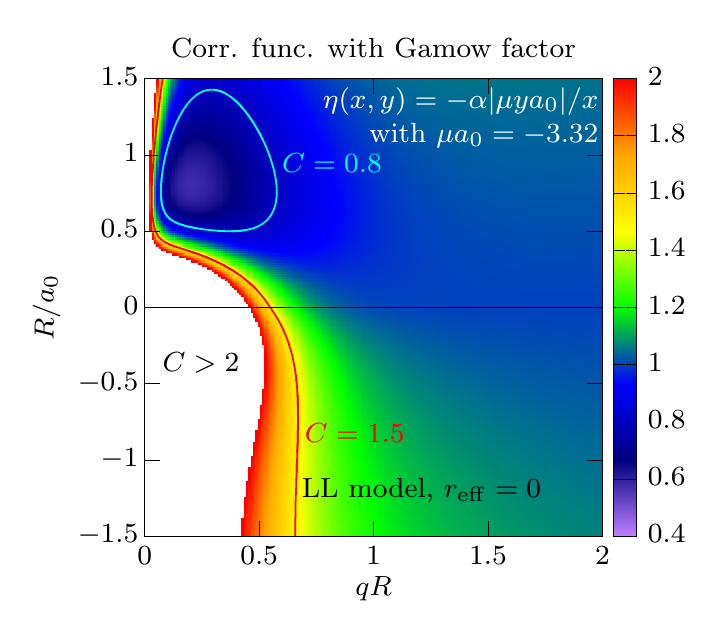}}
	\caption{Same 
as Fig~\ref{fig:LLcorr} but with Gamow factor. For given $(x,y)=(qR, R/a_0)$, $\eta = -\mu\alpha/q$ is calculated as $\eta(x,y) = -\alpha |\mu y a_0|/x$  where  we adopt
$\mu a_0 = \mu_{p\Xi^-} a_0^{N\Xi (J=0)} = -3.32$.}
	\label{fig:LL-C}
\end{center}
\end{figure}

When the Coulomb attraction operates on top of the strong interaction, $C(q)$ is enhanced in the low $q$ region
and the suppression found in Fig.~\ref{fig:LLcorr} with $a_0 > 0$ (without the Coulomb potential) is expected to appear 
as a dip of $C(q)$  when the source size $R$  is comparable to $a_0$.
This is illustrated  in Fig.~\ref{fig:LL-C}
where  the Coulomb effect is considered qualitatively  by multiplying the Gamow factor given as $A_{\mathrm{Gamow}}(\eta) = 2\pi\eta/ (\exp(2 \pi \eta)-1) $. 
 On the other hand, in the negative $a_0$  region without the bound state, the dip structure is not expected in $C(q)$ for wide range of $R=$1-5 fm.
Recent preliminary data from Au+Au collisions~\cite{Mi:2021} seem to show no dip in the $p\Xi^-$ correlation function, which is consistent with the HAL QCD potential where  there is no quasi-bound state of $p\Xi^-$ generated by the strong interaction.

\input{FemHHI-refs.tex}
\end{document}

%% file: FemHHI-refs.tex


%% file: pXi_v4f.bbl
\begin{thebibliography}{10}

\bibitem{Oka:1988yq}	
	M. Oka,
	Phys. Rev. D {\bf 38} (1988), 298.

\bibitem{Gal:2015rev}	
	A. Gal,
	Acta Phys. Polon. B {\bf 47} (2016), 471
	[arXiv 1511.06605 [nucl-th]].

\bibitem{Clement:2016vnl}	
	H. Clement,
	Prog. Part. Nucl. Phys. {\bf 93} (2017), 195
	[arXiv 1610.05591 [nucl-ex]].

\bibitem{Jaffe:1976yi}	
	R. L. Jaffe,
	Phys. Rev. Lett. {\bf 38} (1977), 195 [Erratum: Phys. Rev. Lett.38,617(1977)].

\bibitem{Takahashi:2001nm}	
	H. Takahashi et al.,
	Phys. Rev. Lett. {\bf 87} (2001), 212502.

\bibitem{Nakazawa:2010zzb}	
	K. Nakazawa and H. Takahashi,
	Prog. Theor. Phys. Suppl. {\bf 185} (2010), 335.

\bibitem{Morita:2014kza}	
	K. Morita, T. Furumoto, and A. Ohnishi,
	Phys. Rev. C {\bf 91} (2015), 024916
	[arXiv 1408.6682].

\bibitem{Acharya:2018gyz}	
	S. Acharya et al. [ALICE],
	Phys. Rev. C {\bf 99} (2019), 024001
	[arXiv 1805.12455].

\bibitem{Acharya:2019yvb}	
	S. Acharya et al. [ALICE],
	Phys. Lett. B {\bf 797} (2019), 134822
	[arXiv 1905.07209].

\bibitem{Sasaki:2019qnh}	
K. Sasaki et al. [HAL QCD],
Nucl. Phys. A {\bf 998} (2020), 121737
[arXiv 1912.08630].


\bibitem{Nakazawa:2015joa}	
	K. Nakazawa et al.,
	PTEP {\bf 2015} (2015), 033D02.

\bibitem{Hayakawa:2020oam}
S. H. Hayakawa \textit{et al.} [J-PARC E07],
Phys. Rev. Lett. \textbf{126} (2021),  062501

\bibitem{Yoshimoto:2021ljs}
M.~Yoshimoto \textit{et al.} [J-PARC E07],
Prog. Theor. Exp. Phys. {\bf 2021} (2021), 073D02



\bibitem{Acharya:2019sms}	
	S. Acharya et al. [ALICE],
	Phys. Rev. Lett. {\bf 123} (2019), 112002
	[arXiv 1904.12198].
\bibitem{Acharya:2020asf}
	S. Acharya et al. [ALICE],
	Nature {\bf 588} (2020), 232-238
	[arXiv 2005.11495].

\bibitem{Haidenbauer:2015zqb}	
	J. Haidenbauer, U. G. Mei\ss{}ner, and S. Petschauer,
	Nucl. Phys. A {\bf 954} (2016), 273-293
	[arXiv 1511.05859 [nucl-th]].



\bibitem{Li:2018tbt}
         K. W. Li, T. Hyodo, and L. S. Geng,
	Phys. Rev. C \textbf{98}  (2018), 065203 

\bibitem{Petschauer:2020}
	S. Petschauer, J. Haidenbauer, N. Kaiser, Ulf-G. Meissner, and W. Weise,
	Frontiers in Physics {\bf 8} (2020), 146.

\bibitem{HALQCD:2020}
       S. Aoki and T. Doi, Frontiers in Physics {\bf 8} (2020), 307.


\bibitem{Nishizaki:2002ih}
S.~Nishizaki, T.~Takatsuka, and Y.~Yamamoto,
Prog. Theor. Phys. {\bf 108} (2002), 703-718

\bibitem{Lattimer:2021}
J.M. Lattimer, 
 Ann. Rev. Nucl.  Part. Sci. {\bf  71} (2021), 433-64.

\bibitem{Ohnishi:2016elb}	
	A. Ohnishi, K. Morita, K. Miyahara, and T. Hyodo,
	Nucl. Phys. A {\bf 954} (2016), 294-307
	[arXiv 1603.05761].

\bibitem{Morita:2016auo}
	K. Morita, A. Ohnishi, F. Etminan, and T. Hatsuda,
	Phys. Rev. C {\bf 94} (2016), 031901 [Erratum: Phys. Rev. C 100, 069902 (2019)]
	[arXiv 1605.06765].

\bibitem{Hatsuda:2017uxk}	
	T. Hatsuda, K. Morita, A. Ohnishi, and K. Sasaki,
	Nucl. Phys. A {\bf 967} (2017), 856-859
	[arXiv 1704.05225].

\bibitem{Mihaylov:2018rva}	
	D. L. Mihaylov, V. Mantovani Sarti, O. W. Arnold, L. Fabbietti, B. Hohlweger, and A. M. Mathis,
	Eur. Phys. J. {\bf C78} (2018), 394
	[arXiv 1802.08481].

\bibitem{Haidenbauer:2018jvl}	
	J. Haidenbauer,
	Nucl. Phys.  A {\bf 981} (2019), 1
	[arXiv 1808.05049].

\bibitem{Morita:2019rph}	
	K. Morita, S. Gongyo, T. Hatsuda, T. Hyodo, Y. Kamiya, and A. Ohnishi,
	Phys. Rev. C {\bf 101} (2020), 015201
	[arXiv 1908.05414].

\bibitem{Kamiya:2019uiw}	
	Y. Kamiya, T. Hyodo, K. Morita, A. Ohnishi, and W. Weise,
	Phys. Rev. Lett. {\bf 124} (2020), 132501 (1-6)
	[arXiv 1911.01041].

\bibitem{Fabbietti:2020bfg}
L.~Fabbietti, V.~M.~Sarti, and O.~V.~Doce,
[arXiv:2012.09806 [nucl-ex]].

\bibitem{Koonin:1977fh}	
	S. E. Koonin,
	Phys. Lett. B {\bf 70} (1977), 43.

\bibitem{Pratt:1986cc}
	S. Pratt,
	Phys. Rev. D {\bf 33} (1986), 1314.

\bibitem{Anchishkin:1997tb}	
	D. Anchishkin, U. W. Heinz, and P. Renk,
	Phys. Rev. C {\bf 57} (1998), 1428-1439
	[arXiv nucl-th/9710051 [nucl-th]].

\bibitem{Lednicky:1981su}	
	R. Lednicky and V. L. Lyuboshits,
	Sov. J. Nucl. Phys. {\bf 35} (1982), 770 [Yad. Fiz.35,1316(1981)].

\bibitem{Lednicky:1998r}	
	R. Lednicky, V. V. Lyuboshits, and V. L. Lyuboshits,
	Phys. Atom. Nucl. {\bf 61} (1998), 2950.

\bibitem{Ishii:2006ec}	
N. Ishii, S. Aoki, and T. Hatsuda,
Phys. Rev. Lett. {\bf 99} (2007), 022001
[arXiv nucl-th/0611096].

\bibitem{HALQCD:2012aa}	
N. Ishii et al. [HAL QCD],
Phys. Lett. B {\bf 712} (2012), 437
[arXiv 1203.3642].


\bibitem{Seaton:2002}
	M. J. Seaton,
	Comp. Phys. Comm. {\bf 146} (2002), 225.


\bibitem{Noble:2004eva}
	C. Noble,
	Comp. Phys. Comm. {\bf 159} (2004), 55.


\bibitem{Acharya:2019bsa}
	S. Acharya et al. [ALICE],
	Phys. Rev. Lett. {\bf 124} (2020), 092301
	[arXiv 1905.13470].


\bibitem{Borsanyi:2013bia}
	S. Borsanyi et al.,
	Phys. Lett. B {\bf 730} (2014), 99
	[arXiv 1309.5258].

\bibitem{Bazavov:2014pvz}
	A. Bazavov et al. [HotQCD],
	Phys. Rev. D {\bf 90} (2014), 094503
	[arXiv 1407.6387].


\bibitem{Adamczyk:2014vca}
	L. Adamczyk et al. [STAR],
	Phys. Rev. Lett. {\bf 114} (2015), 022301
	[arXiv 1408.4360].


\bibitem{Ohnishi:2021ger}
A. Ohnishi, Y. Kamiya, K. Sasaki, T. Fukui, T. Hatsuda, T. Hyodo, K. Morita, and K. Ogata,
Few Body Syst. \textbf{62}, 42 (2021).


\bibitem{Mrowczynski:2019yrr}
	S. Mr\'owczy\'nski and P. S\l{}o\'n,
	Acta Phys. Polon. B {\bf 51} (2020), 1739-1755
	[arXiv 1904.08320 [nucl-th]].

\bibitem{Etminan:2019pru}
	F. Etminan and M. M. Firoozabadi,
arXiv 1908.11484 [nucl-th].

\bibitem{Haidenbauer:2020uew}
	J. Haidenbauer,
	Phys. Rev. C {\bf 102} (2020), 034001
	[arXiv 2005.05012 [nucl-th]].

\bibitem{Ogata:2021juh}
	K. Ogata, T. Fukui, Y. Kamiya, and A. Ohnishi,
Phys. Rev. C {\bf103} (2021), 065205
arXiv 2103.00100 [nucl-th].


\bibitem{Ahn:1998fj}	
	J. Ahn et al. [KEK-PS E224],
	Phys. Lett. B {\bf 444} (1998), 267.

\bibitem{Yoon:2007aq}	
	C. J. Yoon et al. [KEK-PS E522],
	Phys. Rev. C {\bf 75} (2007), 022201(R).

\bibitem{Ahn:2017}	
	J. K. Ahn [J-PARC E42],
	J. Phys. Soc. Jpn. Conf. Proc. {\bf 17} (2017),  031004.

\bibitem{Masui:2000mug}	
H. Masui, S. Aoyama, T. Myo, K. Kat\={o}, and K. Ikeda,
Nucl. Phys. A {\bf 673} (2000), 207-218.

\bibitem{Mi:2021}
K. Mi et al.  [STAR], 
talk at APS April Meeting 2021, L13.00007 (2021).



\end{thebibliography}
